%% file: ms.tex
\documentclass[12pt,preprint]{aastex}
\def\msun{\rm M_{\sun}}

\def\LOri{${\lambda}$ Orionis}
\def\av{${\rm A_V}$}
\def\Rc{${\rm R_C}$}
\def\Ic{${\rm I_C}$}

\begin{document}
\shortauthors{Hernandez et al.}
\shorttitle{{\em Spitzer} Observations of the $\lambda$ Orionis cluster. II}

\title{ {\em Spitzer} Observations of the $\lambda$ Orionis cluster. II. Disks around solar-type and low mass stars}

\author{Jes\'{u}s Hern\'andez\altaffilmark{1}, Maria Morales-Calderon\altaffilmark{2}, 
Nuria Calvet\altaffilmark{3}, L. Hartmann\altaffilmark{3}, J. Muzerolle\altaffilmark{4}, 
R. Gutermuth\altaffilmark{5}, K. L. Luhman\altaffilmark{6}, J. Stauffer\altaffilmark{7}}

\altaffiltext{1}{Centro de Investigaciones de Astronom\'{\i}a, Apdo. Postal 264, M\'{e}rida 5101-A, Venezuela}

\altaffiltext{2}{Laboratorio de Astrof´ısica Estelar y Exoplanetas (LAEX), Centro de Astrobiolog´ıa (CAB, INTA-CSIC), 
LAEFF, P.O. 78, E-28691, Villanueva de la canada, Madrid, Spain}

\altaffiltext{3}{Department of Astronomy, University of Michigan, 830 Dennison Building, 500 Church Street, Ann Arbor, MI 48109, USA}

\altaffiltext{4}{Steward Observatory, University of Arizona, 933 North Cherry Avenue, Tucson, AZ 85721, USA}

\altaffiltext{5}{Harvard-Smithsonian Center for Astrophysics, 60 Cambridge, MA 02138, USA}

\altaffiltext{6}{Center for Exoplanets and Habitable Worlds, The Pennsylvania State University, University Park,
PA 16802, USA}

\altaffiltext{7}{{\em Spitzer} Science Center, Caltech M/S 220-6, 1200 East California Boulevard, Pasadena, CA 91125, USA}

\email{hernandj@cida.ve}

\begin{abstract}

We present IRAC/MIPS {\em Spitzer Space Telescope} observations of the solar type and 
the low mass stellar population of the young ($\sim$ 5 Myr) {\LOri} cluster.  
Combining optical and 2MASS photometry, we 
identify 436 stars as probable members of the cluster. 
Given the distance (450 pc) and the age of the cluster, our sample ranges in mass from
2$\msun$ to objects below the substellar limit.
With the addition of the {\em Spitzer}  mid-infrared data, we have identified 49 stars bearing 
disks in the stellar cluster. 
Using spectral energy distribution (SED) slopes, we place objects in several classes:
non-excess stars (diskless), stars with optically thick disks,
stars with ``evolved disks''( with smaller excesses than optically thick disk systems), and 
``transitional disks'' candidates (in which the inner disk is partially or fully cleared).
The disk fraction depends on the stellar mass, ranging from $\sim$6\% for
K type stars (\Rc-J$<$2) to $\sim$27\% for stars with spectral type M5 or later (\Rc-J$>$4). 
We confirm the dependence of disk fraction on stellar mass in this age range found in other studies. 
Regarding clustering levels, the overall fraction of disks in the {\LOri}
cluster is similar to those reported in other stellar groups with ages normally quoted
as $\sim$5 Myr. 

\end{abstract}

\keywords{Stars: formation --- Stars: pre-main sequence --- Infrared: stars
--- Protoplanetary disks --- open clusters and associations: individual (Lambda Orionis Cluster)}

\section{INTRODUCTION}
\label{sec:int}

Circumstellar disks, which appear to be a natural byproduct of the process
of star formation, play a critical role in the evolution of stars and planetary systems. 
As these disks evolve, the rate of accretion onto the central star
decreases and the gas dissipates, while some of the remaining dust is thought to
coalesce into large bodies such as planets and planetesimals. The more crucial
processes in the evolution from primordial disks 
(e.g., optically-thick disks with modest dust coalescence)
to planetary systems occur
in the age range $\sim$ 1 - 10 Myr, when the dust and gas in the disk
is removed \citep{calvet05}, second generation dust is created by
collisional cascade \citep{hernandez06,currie08}, and planets are expected to
form \citep{podosek94}.  
Studies with the {\em Spitzer Space Telescope} \citep{werner04}
of the disk evolution of young, low mass ($\leq$ 2 $M_{\sun}$)  pre-main
sequence stars (T Tauri stars; TTS) in star-forming regions
indicate that at 5 Myr about 80\% of primordial disks have 
dissipated \citep{carpenter06,dahm07,hernandez07a,hernandez08} in agreement with
results using near infrared observations \citep{haisch01,hillenbrand06}.   
The disk infrared emission observed in stars bearing primordial 
disks also declines with age \citep{hernandez07b} suggesting that as time passes, 
dust grains in the  disk collide, stick together, grow to sizes much larger than the
wavelength of observation and settle toward the midplane, which reduces the 
flaring of the disk and thus the amount of energy radiated 
\citep{kh87,dullemond05,dalessio06}. Finally, $\sim$5 Myr 
is the age at which observations \citep[e.g.;][]{hernandez09} and 
models \citep[e.g.][]{kenyon04,kenyon08} suggest
that second generation disks, which may trace active 
planet formation \citep{greenberg78,backman93,kenyon08,wyatt08,cieza08},
start to dominate the disk population around intermediate mass stars. 

The {\LOri} star forming region is one of the most prominent 
OB associations in Orion. It includes 
a ring-like structure of dust and gas of 8-10 degrees in 
diameter \citep[e.g.;][]{maddalena87,zhang89,dolan02}.  Several 
young stellar regions are located near the ring (e.g. B30, B35, LDN1588, LDN1603), 
in which low mass star formation continues today \citep{barrado07a,mathieu08,morales09}. 
The {\LOri} cluster (also known as Collinder 69) is located in a region 
nearly devoid of dense gas near the center of the ring. 
According to several authors \citep{maddalena87, cunha96, dolan99, dolan02}, 
{\LOri} (spectral type O8 III)  had a massive companion that
became a supernovae (SN), removing nearby molecular gas at the center of 
the cluster about 1 Myr ago. The ring of dust 
and gas is then the current location of the swept up material 
from the SN \citep[e.g.;][]{mathieu08,morales09}.

Because the {\LOri} cluster is reasonably nearby \citep[450 pc; see][]{mathieu08}
and relatively populous, 
it represents a valuable laboratory for studies of disk evolution,
making statistically significant studies of disk properties in a wide range of stellar masses. 
Its age of 4-6 Myr \citep[e.g.,][]{murdin77, dolan01} represents 
an intermediate evolutionary stage between two populous stellar clusters 
located at similar distances in the Orion star formation complex \citep{bally08} and studied 
using similar methods to detect and characterize their disks: 
the $\sigma$ Orionis cluster \citep[$\sim$3 Myr;][]{hernandez07a} and 
the 25 Orionis aggregate \citep[7-10 Myr;][]{hernandez07b,briceno07}.

The {\em Spitzer Space Telescope} with its unprecedented sensitivity 
and spatial resolution in the near- and mid-infrared
windows is a powerful tool to expand significantly our understanding
of star and planet formation processes; it provides resolved near- and 
mid-infrared photometry for young stellar populations down to relatively 
low masses. 
In an previous paper \citep[][hereafter Paper I]{hernandez09}, we studied the disk 
population around intermediate mass stars with spectral types F 
and earlier in the {\LOri} cluster. In this contribution, 
we complete a census of the disk population in the cluster studying 
the infrared excesses produced by 
disks  around stars ranging 
in mass from solar type stars to  the substellar limit. 
This paper is organized as follows. In Section \ref{sec:obs}, we describe the
observational data in the cluster. In Section \ref{sec:selmem}, we describe the
selection of possible members.
In Section \ref{sec:res}, we present
the method for identifying stars with infrared excesses and the classification
of the disk population. 
In Section \ref{disk_frec}, we present 
the disk frequencies of the cluster and we compare these frequencies with 
results from other stellar groups. Finally, we present our conclusions 
in Section \ref{sec:conc}. 

\section{OBSERVATIONS}
\label{sec:obs}
\defcitealias{barrado07b}{BN07}
\defcitealias{dolan02}{DM02}
\defcitealias{hernandez09}{Paper I}
\subsection{{\em Spitzer} observations}
\label{sec:ir}

We are using the {\em Spitzer Space Telescope} observations 
from the \citetalias{hernandez09}, which include images from
the InfraRed Array Camera \citep[IRAC,][]{fazio04} 
with its four photometric channels (3.6, 4.5, 5.8 \& 8.0 \micron) 
and from the 24{\micron} band of the Multiband Imaging Spectrometer 
for {\em Spitzer} \citep[MIPS][]{rieke04}. 

The region with the complete IRAC data set (hereafter IRAC region) 
has a size of {$\sim$57.6\arcmin x54.6\arcmin} centered at \LOri.
More than 40,000 sources were detected in at least one IRAC band using 
PhotVis (version 1.09), an IDL GUI-based photometry visualization tool 
developed by R. Gutermuth using the DAOPHOT modules  ported to IDL 
as part of the IDL Astronomy User Library \citep{landsman93}.
Final IRAC photometric errors include the uncertainties in the zero-point 
magnitudes ($\sim$0.02 mag). 
MIPS observations cover more than 99\% of the IRAC region.
The absolute flux calibration uncertainty is $\sim$4\% \citep{engelbracht07}.
Our final flux measurements at 24{\micron} are complete down to about 0.8m Jy.
The detailed description of the observation, reduction and calibration 
of the IRAC and MIPS images is given in \citetalias{hernandez09}. 

The Figure \ref{f:map} illustrates the location of the IRAC region 
(left-panel), overlaid on a map of dust infrared emission \citep{schlegel98} 
and CO isocontours \citep{dame01}. The space distribution of disk 
bearing stars of the {\LOri} cluster (see \S\ref{sec:res}) is shown
on a false color image (right panel).

\subsection{Optical and Near-IR photometry}
\label{sec2:opt}

We primarily used the V{\Rc\Ic} photometry from 
\citet[][ hereafter DM02]{dolan02} 
to identify possible members in the {\LOri} cluster. 
These data were obtained at the Kitt Peak National Observatory
using the Mosaic imager on the
0.9m telescope covering an area of 60 {$\deg^2$} centered 
in the {\LOri} cluster. Most of the objects in this 
catalog have magnitudes ranging from V$\sim$10 to V$\sim$19. 
We augmented the optical data set to include the {\Rc} and {\Ic}  photometry 
of 111 very low mass stars and brown dwarfs studied 
by \citet[][ hereafter BN07]{barrado07b} not cataloged by \citetalias{dolan02}.  
Since there are systematic differences between 
the {\Rc} and {\Ic} magnitudes of \citetalias{dolan02} 
and  \citetalias{barrado07b} catalogs, we converted 
the \citetalias{barrado07b} photometry to the \citetalias{dolan02} photometric 
system. We fitted a straight line in a magnitude-magnitude plot using 
stars included in both catalogs to obtain the conversion factors 
needed for each filter. 

We combined the optical data set with photometry at J, H and K$_S$ 
from the 2MASS Point Source Catalog \citep{skrutskie06}.
Out of 5168 optical sources in the IRAC region,  4966 stars (96.1\%) 
have 2MASS counterparts and 21 additional sources have near infrared 
(NIR) photometry from \citetalias{barrado07b}. 
About 2752 sources in the 2MASS catalog 
located in the IRAC region do not have optical information. 
In general, sources in this sample are below the optical detection limit. 
In appendix \ref{app1}, we present an IRAC/MIPS analysis for this sample 
to search for sources with infrared excesses.

\section{MEMBERSHIP SELECTION}
\label{sec:selmem}

\subsection{Compilation of known members}
\label{sec:known}

The {\LOri} star forming region has been the subject of several 
studies intended to identify members of its stellar and substellar 
population and some of these overlap with our IRAC region. 
\citet{duerr82} surveyed a 
region of $\sim$100 $\deg ^2$ around $\lambda$ Orionis identifying nearly 
100 H$\alpha$ emission objects (they used an objective-prism survey). 
However, most of these objects are located 
outside the central region, 
and only two of these objects are in the IRAC region (\#1624 and \#4407). 
\citet{dolan01,dolan99}
identified 266 pre-main sequence stars in the {\LOri} star forming 
region using the presence of \ion{Li}{1}$\lambda$6708 in absorption and 
radial velocities. They reported a mean radial velocity of 24.5 
km s$^{-1}$ with small dispersion of radial velocities (2.3 km s$^{-1}$). Of the 
64 sources with \ion{Li}{1} in absorption that we have in common with 
the \citet{dolan01,dolan99} catalogs (the remaining sources are located outside of IRAC region), 
4 stars have radial velocities outside of the 3$\sigma$ range 
(17.6 - 31.4 km s$^{-1}$)  that Dolan \& Mathieu used for identifying 
members of the cluster. Since these stars have \ion{Li}{1} in absorption 
(a diagnostic of youth) and binaries can lie off the cluster velocity \citep[e.g.,][]{tobin09},
we placed them as binaries candidates of the {\LOri} cluster. 

Using optical and near infrared photometry and low resolution spectroscopy, 
\citet{barrado04} reported memberships for 170 low mass stars
and brown dwarf candidates belonging to the {\LOri} cluster. 
This study was improved by \citetalias{barrado07b} 
using new deep NIR photometry, IRAC photometry, and additional 
low resolution spectroscopic data. They reported
19 probable non-members, four stars with dubious membership and 147 
bona-fide members of the cluster (hereafter BN07 members), of which 
more than half have spectroscopic measurements of H$\alpha$ and/or spectral 
types.  
All of the \citetalias{barrado07b} members are in the IRAC region. 
We revisited the disk population of these objects
using methods consistent with previous studies of a numbers 
of star forming regions \citep{hernandez07a,hernandez07b,hernandez08},
including a better estimation of the effects of errors,
for better comparison of disk properties as a function of 
age and environment (see \S \ref{sec:res}).

Recently, \citet{maxted08} and \citet{sacco08} studied 
sub-samples of the \citetalias{barrado07b} members using 
high resolution FLAMES \citep[Fiber Large Array Multi Element Spectrograph][]{pasquini02} spectra. 
Based on radial velocity measurements, 
\citet{maxted08} confirmed 69 \citetalias{barrado07b} members and rejected 4 
using a radial velocity range for members of 22-32 km s$^{-1}$. 
Combining three independent criteria
including  the radial velocity distribution, the presence of the
\ion{Li}{1}$\lambda$6708 line and the H$\alpha$ line, \citet{sacco08} reported 
membership for 44 \citetalias{barrado07b} members and 5 stars confirmed
as members of the cluster by \citet{dolan01}. Since \citet{sacco08}
did not detect \ion{Li}{1} in absorption for four \citetalias{barrado07b} 
members, they reported them as non-members of the cluster.

Table \ref{tab:known_mem} summarizes the compiled membership information.
Column (1) shows the internal running identification number in our sample;
columns (2) and (3) provide the stellar coordinates;
column (4) shows other names used previously; columns (5) and (6) show spectral types
and their references; columns (7) and (8) show radial velocity measurements and their 
references, columns (9) and (10) indicate whether the star has \ion{Li}{1} in 
absorption and their references, respectively. 
Column (11) shows membership information compiled from previous studies. 
Based on column (11), we split the sample of Table \ref{tab:known_mem} in 
several groups labeled in column 12: (m1) represents {\it confirmed members} of the 
cluster with \ion{Li}{1} in absorption and radial velocity in the member range; 
(m2) represents stars with radial velocity in the member range but no information 
about \ion{Li}{1} ({\it radial velocity members}); (m3) represents stars 
with \ion{Li}{1} in absorption and radial velocity out of the member 
range ({\it binaries candidates}); (m4) represents stars with spectral 
types and photometric data in agreement with the expected trend in color magnitude diagrams ({\it spectral-type members}); 
(m5) represents known members based only on photometric criteria ({\it photometric known candidates});
(nm1) represents non members based on photometric criteria, and (nm2) represents
non members based on spectroscopic analysis.

\subsection{Photometric selection}
\label{s:photsel}

To select additional candidates to the ones described in \S \ref{sec:known}, 
we  used photometric criteria based  on their optical-2MASS colors and magnitudes. Figure \ref{f:cmd_sel} 
shows several color magnitude diagrams (CMDs) illustrating our photometric selection. 
First, we estimated empirical isochrones using the location 
of {confirmed members} (open circles) and other members selected using 
spectroscopic criteria (e.g. open squares; {radial velocity members}, {binaries candidates},
and {spectral-type members}) on these CMDs (hereafter  {confirmed members},
{radial velocity members},  {binaries candidates} and {spectral-type members}
are named as ``known members'' to distinguish from {photometric known candidates} 
which are not used in this procedure).  
Empirical isochrones (dashed lines) were estimated using the median colors 
(V-J, V-\Ic, \Rc-J \& \Rc-\Ic) of the known members for 1 magnitude bins in the V-band 
(upper panels) or in the \Rc-band (bottom panels). Standard deviations ($\sigma$)
were calculated using the differences between the observed colors of the known members
and the expected colors from the empirical isochrones. In each CMD, we fitted two lines
using a second order polynomial corresponding to the points representing by the median color + 2.5$\sigma$
and by the median color - 2.5$\sigma$, respectively (dotted lines); 
the regions of probable members are defined between these lines.   

Out of 5168 sources with optical photometry within the IRAC region, 
4430 sources ($\sim$86\%) are likely background sources since they are 
located below the regions of probable members in all CMDs.
We also found 24 foreground candidates located above the regions 
of probable members in all CMDs.  In the right panels of Figure \ref{f:cmd_sel}, 
we plotted stars with no 2MASS counterparts (symbol X). Stars brighter than 
the magnitude limit in \citetalias{dolan02} catalog (V$\sim$20, R$\sim$18.7) with no 2MASS counterpart 
are located below the regions of probable members and thus are likely 
background stars (220 sources). Stars fainter than the Dolan's photometric limit 
were taken from \citetalias{barrado07b},  in which most of them have NIR photometry 
from deeper images. All of \citetalias{barrado07b} members are located in the regions 
of probable members in the lower panels of Figure \ref{f:cmd_sel} 
(there are not V-band magnitudes for the faintest BN07 members).

Our final sample of candidates of the {\LOri} cluster 
includes 340 stars located in the regions of probable members in at least three of the four CMDs 
(297 are located in the four member regions). Additionally, we included 
96 stars from \citetalias{barrado07b} located in the member regions in the 
lower panels of Figure \ref{f:cmd_sel}. This sample of 436 stars (hereafter $\lambda$Ori sample)
includes 150 known members and 32 photometric known candidates.
Since a giant branch crosses the empirical isochrones at V-J$\sim$2.5, V-\Ic$\sim$1.5, \Rc-J$\sim$1.8 and 
\Rc-\Ic$\sim$0.7, large contamination by no-members of the cluster are expected
around these colors.

Table 2  shows IRAC and MIPS photometry for the $\lambda$Ori sample.
Column (1) shows the internal running identification number;
column (2) provides the 2MASS object name; columns (3) and (4) provide the 
stellar coordinates; columns (5), (6), (7) and (8) give the IRAC magnitudes
in the bands [3.6], [4.5], [5.8] and [8.0], respectively; column (9) gives 
the flux at 24{\micron} (MIPS band); column (10) shows the disk classification
obtained in \S \ref{sec:res}; column (11) shows references for stars studied
previously. 

Finally, a group of 77 sources are located in the region of probable members 
in one or two CMDs but appear as likely foreground or background sources 
in the other CMDs. Infrared properties of these stars with uncertain 
photometric membership classification are also studied in Appendix \ref{app1}.

\section{DISK POPULATION}
\label{sec:res}
To characterize the disk population in the {\LOri} cluster,
we need to identify stars in the $\lambda$Ori sample (see \S \ref{s:photsel}) 
that exhibit excess emission at the IRAC/MIPS bands. For this purpose, we must 
properly account for uncertainties and biases in the 
IRAC/MIPS photometry, particularly near the detection
limit of these data \citep{luhman08}. In a well-populated sample of 
stars not strongly affected by extinction, the sum 
of all these uncertainties and biases (Poisson errors, 
uncertainties in the zero-point magnitudes, location-dependent 
variations in the calibration, scatter of photospheric colors 
among cluster members) is reflected in the spread of colors 
in a given magnitude among the diskless stars of the cluster,  
which dominate stellar populations older than $\sim$2-3 Myr 
\citep[e.g.,][]{haisch01,hillenbrand06,hernandez07a, hernandez08}.
Since a disk produces greater excess emission above the stellar 
photosphere at longer wavelengths, we selected the bands at 
8{\micron} and 24{\micron} to identify and characterize disks
in the stellar population of the {\LOri} cluster. First, 
we identified disk bearing stars with infrared excesses 
at 24{\micron} and 8{\micron} in \S\ref{mips_exc} 
and \S\ref{irac_exc}, respectively.  
In \S\ref{disk_type} and \S\ref{seds}, we completed our analysis 
and classification of the disk population of the {\LOri} cluster
based on the excess emission levels observed in the IRAC/MIPS bands
and their SEDs.

\subsection{Identifying stars with  24{\micron} excess}
\label{mips_exc}

Figure \ref{f:mips_excess} illustrates 
the procedure to identify stars with 24{\micron} infrared 
excess above the photospheric level \citep[e.g.;][]{gorlova06,gorlova07,
hernandez06,hernandez07a,hernandez07b, hernandez08}.
The upper panel shows the K-[24] color distribution
of stars in the $\lambda$Ori sample detected in the MIPS 24{\micron} image. 
The distribution of stars with K-[24]$<$1.0 can be described by a Gaussian 
centered at 0.15 with $\sigma$=0.18. The 3$\sigma$ boundaries (dotted lines in 
lower panels) represent the photospheric colors; thus stars with 
excesses at 24{\micron} have colors K-[24]$>$0.69. 
Since the {\LOri} cluster has low  reddening \citep[\av$\sim$0.4; ][]{diplas94} the 
R-J color can be used as a proxy of spectral types. 
We display the  location of the spectral type sequence, using the standard R-J colors
from \citet{kh95}.  The NIR colors from \citet{kh95} were converted 
into the 2MASS photometric system  using the transformations from \citet{carpenter01}.
We extended the standard R-J colors toward later spectral types 
using members with spectral types M6 or later reported by \citetalias{barrado07b}. 
The R-J standard colors were estimated averaging the observed 
R-J color \citepalias{barrado07b} in a given bin of spectral types. 
There are 4 candidates bearing disks with small excesses at 24\micron (K-[24]$<$1.5); three of them have
the lower limit of their error bars inside the photospheric region and thus the 24{\micron} 
excess of these objects is not completely reliable. We have marked these objects in Table \ref{t:main}
as stars with uncertain 24{\micron} excess. Two objects have spectral types $\sim$M6 or later and 
are probable brown dwarfs bearing optically thick disks \citepalias{barrado07b}. 
There are three stars bearing disks with photometric spectral type earlier 
than K5. Most stars with 24{\micron} excess ($>$75\%) have excess ratio 
at 24{\micron} \citepalias[E$_{24}$=$10^{0.4*(K-[24]+0.15)}$; e.g.][]{hernandez09}
larger than 22. The lack of M type stars with photospheric K-[24] 
colors reflects the limit magnitude of the 24{\micron} MIPS band.

\subsection{Identifying stars with  8{\micron} excess}
\label{irac_exc}

Figure \ref{f:irac}  shows the IRAC SED slope versus 
the [8.0] magnitude for stars in the $\lambda$Ori sample 
illustrating the procedure to detect
infrared excesses at 8{\micron} \citep[e.g.;][]{lada06, hernandez07a, hernandez07b, hernandez08}. 
The IRAC SED slope ($\alpha$=$dlog[{\lambda}F_\lambda]/dlog[\lambda]$)
is determined from the [3.6]-[8.0] color. 
The error bars in $\alpha$  are calculated propagating the photometric errors at [3.6] and [8.0]. 
To determine the photospheric levels, we used the mean error of 
$\alpha$ ($\bar{\sigma}$) for stars within 0.5 magnitude bins in 
the [8.0] band. The photospheric region (dotted lines) is determined 
using a 3$\bar{\sigma}$ criteria from the median values of $\alpha$ (long dashed lines). 
The width of this region depends on the 
photometric error of the
[8.0] magnitude, which dominates the 
uncertainties in $\alpha$. 
Thus, a single value of $\alpha$ to separate diskless and disk bearing stars 
based on the IRAC color [3.6]-[8.0] is not useful, especially 
near the IRAC detection limit \citep{hernandez07a}.
At the faintest [8.0] magnitudes, the photospheric region is too 
sparsely populated for a reliable measurement of infrared excesses.
For instance, star \#7968 ([8.0]$\sim$14.5) represents the faintest star 
with a probable disk based on 8{\micron} excess. However, since the detections 
at 5.8{\micron} and 8.0{\micron} of this star are background-limited 
and upward fluctuations of the background noise could contaminate these detections, 
the presence of a disk around this star needs additional confirmation.

For comparison, short-dashed lines in Figure \ref{f:irac} 
represent the limits used by \citetalias{barrado07b} to classify disks around stars 
in the {\LOri} cluster based on the following criteria \citep{lada06}: diskless stars ($\alpha <$-2.56),
optically thick disks ($\alpha >$-1.8), and optically thin disks (-2.56$< \alpha <$-1.8). 
The criteria used by \citetalias{barrado07b} is based on the $\alpha$ distribution 
of the disk population in Taurus (1-2 Myr) from \citet{hartmann05}, where most of disk bearing stars 
have $\alpha >$-1.8. The right panel of Figure \ref{f:irac}  shows the $\alpha$ distribution 
of the disk population in Taurus (1-2 Myr) from \citet{luhman10} and supports that 
disk bearing stars with $\alpha<$-1.8 are relatively scarce in Taurus and more 
frequently observed in older stellar regions \citep{lada06,hernandez07a,hernandez07b,hernandez08}.

In Figure \ref{f:irac}, we also display stars with 24{\micron} excess (open circles) 
detected in \S\ref{mips_exc}. In general, stars brighter than [8.0]$\sim$12.0 with 
significant excesses at 8{\micron} have excesses at 24{\micron}. 
There are some disk bearing candidates with error bars inside the photospheric region, 
the 8{\micron} excess of these objects is uncertain and are identified in Table \ref{t:main} 
(see also \S \ref{seds}). Finally, star \#1039 is a relatively bright 
source ([8.0]$\sim$9.5)  that shows infrared excess only at 8{\micron} 
($>$ 5$\sigma$ above the photospheric limit). Visual inspection 
of the image does not reveal any obvious source of contamination at 8{\micron}. 
However, at 24{\micron} (where the disk produces greater excess emission) 
the star \#1039 exhibits photospheric fluxes and thus the presence of a disk 
around this object is highly uncertain.

\subsection{Disk diagnostics}
\label{disk_type}

Figure \ref{f:diskclass} displays the distribution of disk bearing stars (with MIPS counterparts)
of the $\lambda$Ori sample
in a SED slope diagram, generated using the K-[5.8] and [8.0]-[24] colors. The K-[5.8] slope
represents disk emission at 5.8{\micron}, while the [8.0]-[24] slope indicates
whether the star has a flat (slope$\sim$0) or rising (slope$>$0) SED  at wavelengths greater than 8{\micron}.
Using colors from \citet{luhman10}, we display the SED slope distributions (right-panel and upper panel) 
and the quartiles (large error bars) of the disk population in Taurus.
The dashed line represents a limit where the inner disk 
emission has not been affected significantly by evolutionary processes
\citep[a lower limit of primordial disks for Taurus;][]{luhman10}. About
97\% of the disk bearing stars in Taurus are located above this line. 
We use this boundary as a proxy to separate stars bearing optically thick disks
(above the dashed line) from other types of disks. 
We show in Figure \ref{f:diskclass} the location of five transitional disks in 
several star forming regions: TW Hya, GM Aur, Coku Tau/4, CS Cha and CVSO 224 
\citep{calvet02,calvet05,dalessio05, espaillat07a, espaillat08a}. 
Observationally,  they are characterized by relatively small excesses at 5.8{\micron} and large 
excesses at longer wavelengths. 
Star \#4111 can be identified as a {\it transitional disk} candidate.
Since star \#1152 has a rising SED after 8{\micron} and it is near the 
optically thick disk limit, we also place this star as a 
{\it transitional disk} candidate.
The disk of the star Coku Tau/4 is circumbinary, and its structure 
is a result of tidal truncation due to the binary orbit \citep{ireland08}. There 
is a possibility that our transitional disk candidates can be binary systems 
and their disks are similar to the disk associated with Coku Tau/4.
In addition, Figure \ref{f:diskclass} shows 
the location of the newly discovered class of  pre-transitional disks
(LkCa 15 and UX Tau A), in which gaps in primordial disks rather than holes have been identified 
\citep{espaillat07b,espaillat08b,brown08}. Two objects (\#4021 and \#6866) have similar [8.0]-[24] color than the 
pre-transitional disk  LkCa 15, but with less infrared excesses at 5.8\micron. We placed 
these objects as possible {\it pre-transitional disks} in the cluster; however, a more 
detailed study is necessary to reveal the actual nature of their disks.

About 30\% of the late type stars bearing disks in the {\LOri} cluster exhibit relatively small 
IRAC/MIPS excesses (below the dashed lines and with SED slope [8.0]-[24] $<$ 0) indicating that 
evolutionary  processes (e.g. grain growth, dust sedimentation and/or settling) have been at work,
generating flatter disk structures \citep{dalessio06,dullemond07,manoj10}. 
We identify these sources as {\it evolved disks} stars in Table \ref{t:main}.  
We plotted with squares the three earliest stars with infrared excesses 
(photometric spectral type earlier than K5). One of these stars (\#4155)  
exhibits modest infrared excess at 5.8{\micron} and 24\micron,
while the other two stars (\#3785 and \#7402) exhibit marginal infrared excesses at 24{\micron}
and no excesses at 5.8\micron. For comparison, 
we show the locations of the intermediate mass stars bearing disks studied 
in \citetalias{hernandez09}; except for the Herbig Ae/Be star HD 245185, these 
are debris disks candidates, with no excesses in the IRAC bands and  
varying degrees of excesses at 24\micron.

There are 18 disk bearing candidates with excesses at 8{\micron} but 
without MIPS counterparts (see Figure \ref{f:irac}). 
In general, these stars are below the MIPS detection limit. 
Only star \#3746 could be above the MIPS detection limit; however, 
it is located near the star $\lambda$ Ori which may mask the detection 
at 24{\micron}. The disks around two stars (\#1039 and \#7968) are 
highly uncertain (see \S \ref{irac_exc}). 
In summary, based only on the IRAC SED slopes (\S\ref{irac_exc}), 
we found 6 stars with optically thick disks, 7 evolved disk stars and 4 stars  with 
uncertain 8{\micron} excess emission. SEDs of individual objects will be detailed 
in the next section.

\subsection{Spectral Energy Distributions}
\label{seds}

Figures \ref{f:Ltype1} and \ref{f:Ltype2} show SEDs of 
the candidates bearing disks with and without MIPS 
counterparts, respectively. 
Fluxes calculated from optical magnitudes \citepalias{barrado07b,dolan02},
2MASS photometry (J, H \and K), IRAC bands (3.6, 4.5, 5.8 and 8.0 \micron) 
and the 24{\micron} MIPS band were normalized to the flux in the J band.
We sorted the SEDs by photometric spectral types, which where calculated  
interpolating the R-J color in the standard color sequence \citep{kh95}. 
Since the R-J colors are not corrected by reddening, the photometric 
spectral type corresponds to a later limit. 
We display the median SED of the disk population in Taurus. 
It was constructed from the median colors (2MASS, IRAC and MIPS) 
of low mass stars (with spectral types K and M) bearing disks from
\citet{luhman10}. We also show the photospheric fluxes for a star with 
the corresponding photometric spectral type \citep{kh95} normalized 
to the J band. 
To compare our results with those from \citetalias{barrado07b},
we display the IRAC/MIPS photometry given by \citetalias{barrado07b},
with their identification number, disk type and spectral type 
(when is available). In general, disk types identified in 
\S \ref{disk_type} agree with those given by \citetalias{barrado07b}.
Star \#1152 \citepalias[Spectral type M2; thick disk in][]{barrado07b} shows small 
or no infrared excesses at wavelengths shorter than 8{\micron} 
and infrared excess at 24{\micron} comparable to stars 
bearing an optically thick disk; we classified this star
as a transitional disk candidate. 
On the other hand, the star \#3597 \citepalias[Spectral type M4; transition disk in][]{barrado07b}
exhibits a decreasing SEDs after 8{\micron}. We classified this object as 
an evolved disk star. 

Three disk bearing stars with photometric spectral types K7 or earlier 
(\#3785 \#7402 and \#1310) have marginal infrared excesses at 24{\micron} 
and no excesses detected in the IRAC bands. The disk suggested for these objects 
needs additional confirmation. 
If disks are present around these stars, 
their infrared excesses are  comparable to those observed in the early 
type debris disk candidates \citepalias{hernandez09} indicating probably a 
similar origin of their infrared emission \citep[e.g][]{gorlova07,kenyon05}. 
The K-type star \#4155 clearly exhibits infrared excesses at 8{\micron} 
and moderate excess at 24{\micron}. The excesses around 
this star could be explained if the disk is in an  intermediate phase 
evolving from a primordial to a debris disk (evolved disk). 
The star \#5447 was not included in Figure \ref{f:Ltype2}, the SED of 
this thick disk object is similar to that observed in the star \#7490 
(spectral type M4).

In Figure \ref{f:Ltype2}, there are three stars 
(\#7517, \#3710 \& \#2712 with spectral types M0, M6 and M7, respectively) 
with marginal infrared excesses at 8{\micron} which could be produced by either  disk 
emission or PAH (polycyclic aromatic hydrocarbon) contamination. The IRAC photometry of the star \#2712 
could be also contaminated by a near bright source (located at $\sim$7\arcsec) .
Based on the IRAC SED slope (Figure \ref{f:irac}), the star \#7957 is 
classified as an evolved disk star with a marginal excess at 8{\micron}. However, 
the SED observed for this source shows excesses in all IRAC bands similar to 
those observed in sources with thick disks in agreement with the disk type reported by
\citetalias{barrado07b}. 

There are 16 stars with a thin disk type in \citetalias{barrado07b}
that we classified as diskless star based on the criteria discussed above. 
Figure \ref{f:Ltype3} shows SEDs for these objects. 
Several possibilities may support our diskless classification. 
Three stars only have photometry in the [3.6], [4.5] and [5.8] bands in \citetalias{barrado07b}.
Based on the IRAC SED slope, \citetalias{barrado07b} classified as disk bearing stars  
objects with $\alpha >$ -2.56, six of these objects are located in the diskless 
region in the [3.6]-[4.5] versus [5.8]-[8.0] diagram presented by these authors.
As discussed in \S \ref{irac_exc}, the photospheric region of the IRAC slope 
depends on the brightness of the star and thus the single value of the IRAC SED 
slope applied by \citetalias{barrado07b} to separate diskless and disk bearing 
stars does not work well, particularly, for the faintest stars. 
In addition, [8.0] magnitudes
given by \citetalias{barrado07b} are systematically 0.1 magnitudes brighter than our 
photometry. The K-[8.0] colors of the diskless stars 
(with photometric spectral types M5 or earlier) 
in this work are in good agreement with the standard K-[8.0] colors derived from STAR-PET tool available 
on the {\em Spitzer} Science Center website.

\section{DISK FREQUENCIES AND DISK DIVERSITY OF THE {\LOri} CLUSTER}
\label{disk_frec}

The upper panel in Figure \ref{f:disk_frec} shows disk 
frequencies as a function of photometric spectral types estimated using
the \Rc-J color. To reduce the contamination by non-members, 
frequencies were calculated using the members confirmed by \citet{dolan01}, 
\citetalias{barrado07b}, \citet{maxted08} and \citet{sacco08}; see \S{\ref{sec:known}.
We plotted the fractions of confirmed members with optically thick disks 
(filled circles and solid line) and the fractions of confirmed members bearing 
any type of disks (open squares and dotted line) including evolved 
disks and pre-transitional/transitional disks. 
These fractions decrease for decreasing \Rc-J color and flatten up 
around \Rc-J$\sim$4.5 (spectral types M6-M7). The disk fraction may
decline toward later spectral types; however, including the error 
bars, the data are consistent with 
no change in disk frequency from M6 to later types

For reference, we plot the masses 
corresponding to the  \Rc-J colors along the 5 Myr isochrone of \cite{sf00}.
Since \Rc-J color bins represent different 
ranges in stellar masses, the decrease of disk frequencies toward 
earlier spectral types indicates that  primordial disks dissipate 
faster as the stellar mass increases. Particularly, 
for solar type stars the disk frequency is $\sim$6\%, 
while for very low mass stars and brown dwarf mass range 
($M/\msun < $0.1) 
the disk frequency increases to $\sim$27\%. 
The decrease of disk frequencies toward higher stellar mass
agrees with results in other young stellar populations, e.g.: 
IC 348 \citep{lada06}, Upper Scorpius \citep{carpenter06}, 
the {$\sigma$} Ori cluster \citep{hernandez07a}, and the Ori OB1b 
subassociation \citep{hernandez07b}.  For reference, the lower panel of 
Figure \ref{f:disk_frec} shows the location on the color magnitude diagram 
of stars bearing disks selected in \ref{s:photsel}. In general, 
the disk population of stars with spectral type M2 or earlier includes
evolved disks objects while optically thick disks are more frequently 
observed at later spectral types.  

Using the lower panel of Figure \ref{f:disk_frec}, we can estimate the contamination 
levels in our sample. We calculated disk fractions in several \Rc-J color bins
for the confirmed members of the {\LOri} cluster ($F_{disk}^{mem}$) and for 
the entire $\lambda$Ori sample ($F_{disk}^{phot}$). 
Assuming that the source of contamination are photometric candidates 
without infrared excesses and stars with infrared excesses are members 
of the cluster bearing disk,  we can estimate the contamination 
level ($F_{nomem}$) expected in the $\lambda$Ori sample applying
the relation: $F_{nomem}=1-F_{disk}^{phot}/F_{disk}^{mem}$.
Figure \ref{f:contamina} shows the contamination level versus the color \Rc-J.
As discussed in \S\ref{s:photsel}, the larger contamination level is produced by
main sequence stars and by a giant branch that cross the {\LOri} young 
stellar population at \Rc-J$\lesssim$2.5.
In general, the level of contamination for M type stars is less than 25\%.

Regardless of different environments and clustering levels, 
the overall disk frequency for M type stars in the {\LOri} 
cluster (18.5$\pm$4.0 \%) is comparable to those reported in other
stellar groups with ages normally quoted as $\sim$5 Myr.
The more dispersed stellar groups, Upper Scorpius \citep[19$\pm$4 \%, ][]{carpenter06} 
and the Orion OB1b subassociation \citep[15$\pm$4 \%, ][]{hernandez07b}, have
disk frequencies similar to that observed in the {\LOri} cluster. 
This suggests that the clustering levels of stellar groups does not affect significantly 
the disk frequencies observed at 5 Myr. 
In particular, the Orion OB1b subassociation and the {\LOri} cluster have 
similar stellar mass - disk frequency dependencies \citepalias[see][]{hernandez09}.
The disk frequency for stars later than K5  (stellar mass $<$ 1$\msun$)
in the stellar cluster NGC2362 \citep[5 Myr;][]{dahm07} is 
slightly higher (23.0$\pm$3.4) than that for the {\LOri} cluster.
The number of evolved disks may be overestimated by 
\citet{dahm07} as they did not correct for larger photometric errors at faint 
magnitudes (see \ref{sec:res}).
Finally, if the supernovae hypothesis proposed by \citet{dolan99,dolan02} is correct,
the similarities in the disk frequencies for several stellar groups suggest 
that the disk population in the {\LOri} cluster was not affected significantly 
by the supernovae.
 
The detection of evolved disks and pre-transitional/transitional disk 
candidates in the cluster supports the existence of at least two different  
pathways in the evolution from primordial disks to second generation disks 
\citep[e.g.,][]{hernandez07a,cieza08,manoj10,currie10}.
Using the M-type stars with MIPS counterparts (Figures \ref{f:diskclass} and \ref{f:Ltype1}), 
the frequency of pre-transitional/transitional disks in the {\LOri} cluster is 13.3$\pm$6.7 (6.7$\pm$4.7 for the transitional disk candidates alone); 
comparable to that observed in other stellar clusters with ages of 3 Myr or older \citep[see ][]{muzerolle10}. 
The relative scarcity of transitional disks can be used to indicate that 
the phase from primordial disks to second generation disks through 
``the transitional disk evolutionary pathway'' is relatively short. 
While \citet{aurora08} reported a relatively high fraction (40-50\%) of transitional disks 
for the Coronet cluster (age$\sim$1Myr), \citet{ercolano09} 
show that this fraction was overestimated, reporting a transitional 
disk fraction of $\sim$15$\pm$10\% for the cluster. 
On the other hand, one third (33.3$\pm$10.5\%) of the disk population in 
the {\LOri} cluster is classified as evolved or pre-transitional/transitional 
disks (hereafter evolved-transitional disks). 

Using the same method to classify disk populations, the evolved-transitional disk fractions in the 
$\sigma$ Orionis cluster ($\sim$3 Myr), the Orion OB1b subassociation ($\sim$5 Myr) and in the 
25 Orionis (8-10 Myr) aggregate are 17.5$\pm$5.3\%, 35.3$\pm$14.4\% and 40.0$\pm$20.0\%, respectively. 
In contrast, \citet{currie09} reported as many as 80\% of evolved-transitional disk fraction 
for the 5 Myr old cluster NGC2362. Based on this, they reported that the timescale for 
disk dissipation is comparable with the median primordial disk lifetime. 
However, the reported fractions of evolved disks strongly depend on the 
limit used to separate primordial and evolved disks.  
For instance, \citet{currie09} used the lower quartile of the median 
Taurus SED \citep{furlan06} to classify primordial and evolved 
disks ($\sim$25\% of the disk bearing stars in Taurus are located below this limit).  
In Figure \ref{f:diskclass}, we used the boundaries defined by \citet{luhman10}
which follow the upper edges of the color gaps between diskless stars and stars
bearing disks in Taurus. Using this criteria, 97\% of the disk bearing stars 
in Taurus are classified as primordial disks (statistically this means a 
limit of $>$2$\sigma$). Moreover, the disk population in Taurus shows 
a large dispersion in disk emission \citep{hartmann05, luhman10, furlan06, furlan09}.
This could be a reflection of the initial properties of these disks,
of different rates for disk dispersion processes,
and/or of different intrinsic properties of the star-disk systems 
\citep[e.g., multiplicity,  mass of the central object, 
inclination angle of the disk;][]{manoj10, hillenbrand08}. 
Thus, it is difficult to define a reliable criteria to indicate the starting point
in the evolution from primordial to second generation disks and, without an individual 
detailed modeling, the evolved-transitional disk fractions are 
reference values that can be used to compare different disk populations 
assuming an certain criteria. 
In any event, the observation of more evolved-transitional disks stars (given a certain criteria) 
in older stellar groups
does not necessarily imply longer timescales for evolved-transitional disks. 
An alternative is that star formation has stopped and there is no steady 
supply of primordial disks \citep{manoj10}. There is also 
the possibility that the criteria to define primordial disks 
may be limited and some misclassified evolved disks
are still in a primordial disks phase \citep{luhman10,muzerolle10,manoj10}.

\section{SUMMARY AND CONCLUSIONS}
\label{sec:conc}

We have used the IRAC and MIPS instruments on board the {\em Spitzer Space Telescope}
to study the disks population of the {\LOri} cluster. 
Using optical photometry and members confirmed by spectroscopy, 
436 candidates were selected from optical-2MASS color magnitude diagrams. 
The level of contamination by non-members depends on the spectral type range, 
showing the lowest level of contamination ($\lesssim$25\%) for M type stars. 
Combining optical, 2MASS and {\em Spitzer} data and following the procedure 
used in previous disk census \citep{hernandez07a,hernandez07b,hernandez08}, 
we have reported 49 stars bearing disks
, 16 of which were detected only in the IRAC bands.
Out of 49 disk bearing stars, 38 were reported previously as members of the 
{\LOri} cluster \citep[BN07,][]{dolan01}, 11 stars are new 
members of the cluster based on its infrared excesses and its 
optical colors and magnitudes. 

Based on their SEDs and a SED slope diagram, we classified the 
disk bearing stars with MIPS counterpart in three classes:  
20 thick disk stars, nine evolved disk stars, two transitional 
disk candidates and two pre-transitional disk candidates. We found that
for stars with color R-J$>$3 ($\sim$later than M3.5) 
the disk frequencies are 20\%-30\%, 
while stars with colors R-J$<$3 ($\sim$earlier than M3.5) 
the disk frequencies are $<$20\%. 
This indicates a mass dependent timescale for disk dissipation in the {\LOri} cluster, 
similar to results in other young star populations  \citep{carpenter06,hernandez07a,hernandez07b,hernandez08,lada06}.
The overall disk frequency for M type stars in the {\LOri} cluster (18.5$\pm$4.0) is 
similar to those reported in stellar groups with similar evolutionary stage: 
the cluster NGC 2362 \citep[$\lesssim$23$\pm$3, ][]{dahm07},  
the OB association Upper Scorpius \citep[19$\pm$4, ][]{carpenter06} 
and the Orion OB1b subassociation \citep[15$\pm$4, ][]{hernandez07b}. 
This suggests that the clustering level of the stellar group does not 
affect significantly the disk frequencies observed at 5 Myr. 
This work combined with \citetalias{hernandez09} represents a complete disk census of 
the stellar population in the {\LOri} cluster.

\acknowledgements
We thank Fred Adams for his valuable comments and suggestions.
This publication makes use of data products from Two Micron All Sky Survey, 
which is a joint project of the University of Massachusetts and the Infrared 
Processing and Analysis Center/California Institute of Technology. 
This work is based on observations made with the {\em Spitzer Space Telescope} (GO-1 0037), 
which is operated by the Jet Propulsion Laboratory, California Institute of 
Technology under a contract with NASA.  J.H. and N.C gratefully acknowledge 
support from the NASA Origins program grant NNX08AH94G, and the {\em Spitzer} 
General Observer program grant GO1377380.
K. L. was supported by grant AST-0544588 from the National Science Foundation.
The Center for Exoplanets and Habitable Worlds is supported by the
Pennsylvania State University, the Eberly College of Science, and the
Pennsylvania Space Grant Consortium.

\appendix
\section{Additional infrared excess sources}
\label{app1}

We looked for additional members with infrared excesses of the {\LOri} cluster 
analyzing the infrared properties of two samples not included in our 
photometric selection (\S\ref{s:photsel}); stars with 2MASS photometry without 
optical information (\S\ref{sec2:opt}) and stars with uncertain photometric membership
derived from the CMDs (\S\ref{s:photsel}). Out of 2752 2MASS sources in the IRAC region 
without optical information, 43 sources have detections at 24{\micron}. Out of 77 
sources with uncertain photometric membership, 18 sources have detections at 24{\micron}.
Following the procedure described in \S\ref{mips_exc} and \S\ref{irac_exc},
Figure \ref{f:disk_add} shows the diagrams used to identify stars with excesses 
at 24{\micron} and 8{\micron} for these 61 sources with MIPS counterparts.
In the upper panel, we have identified 41 sources with excesses at 24{\micron}. 
Those sources that do not exhibit excesses at 24{\micron} do not exhibit excesses 
at 8{\micron} (bottom panel). On the other hand, there are two sources 
(\#1959 and \#4244) that do not exhibit excesses at 8{\micron} but exhibit excesses at 24\micron. 
In general, sources with IR excesses are 2MASS objects without optical information, 
only one (\#5968) has uncertain photometric membership.
This star is located in the regions of probable members in the V-J versus V 
and the \Rc-J versus {\Rc} diagrams but it appears slightly below from 
the regions of probable members in the V-{\Ic} versus V and the \Rc-{\Ic} 
versus {\Rc} diagrams.

Visual inspection on the IRAC/MIPS images reveals that most sources 
with the largest IR excesses are non-stellar. We found 
that 16 sources are galaxies, and 7 sources have  slightly larger 
PSF (point spread function) widths than other stellar sources 
on the images (labeled as Stellar?). 

Table \ref{t:append} shows IRAC/MIPS photometry 
for the 41 sources with infrared excesses. Column (1) shows the internal running 
identification number; column (2) provides the 2MASS object name; columns (3) 
and (4) provide the stellar coordinates; columns (5), (6), (7) and (8) give 
the IRAC magnitudes in the bands [3.6], [4.5], [5.8] and [8.0], respectively; 
column (9) gives  the flux at 24{\micron} (MIPS band); column (10) shows comments 
based mainly on the visual inspection on the IRAC/MIPS images.
Special notes about some interesting objects are given below Table \ref{t:append}.

\clearpage

\include{tab1s}

\include{tab2s}

\include{tab3}

\clearpage

\begin{figure}
    \centering
    \includegraphics[width=6.0cm, angle=0]{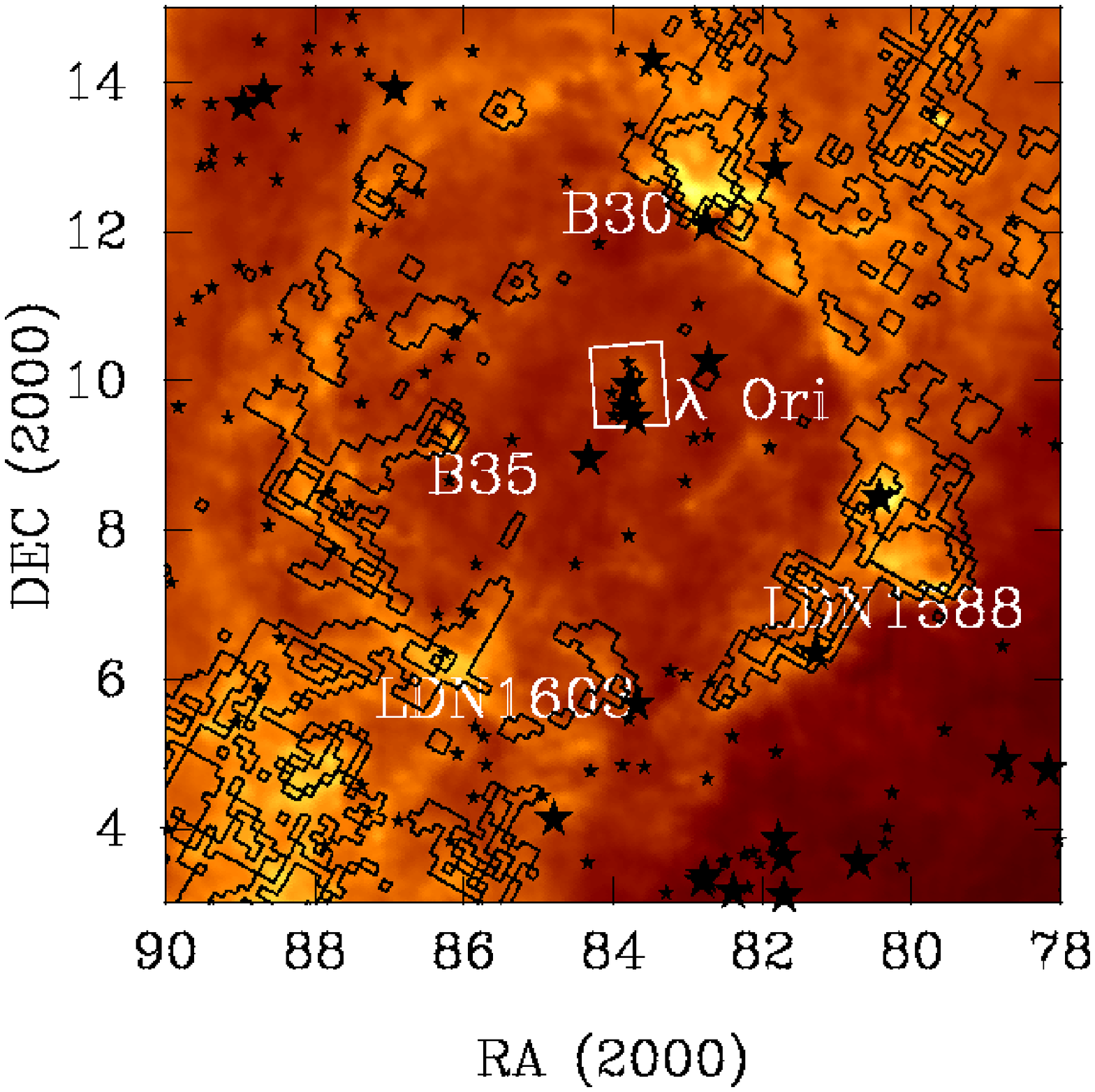}
    \includegraphics[width=8.0cm, angle=0]{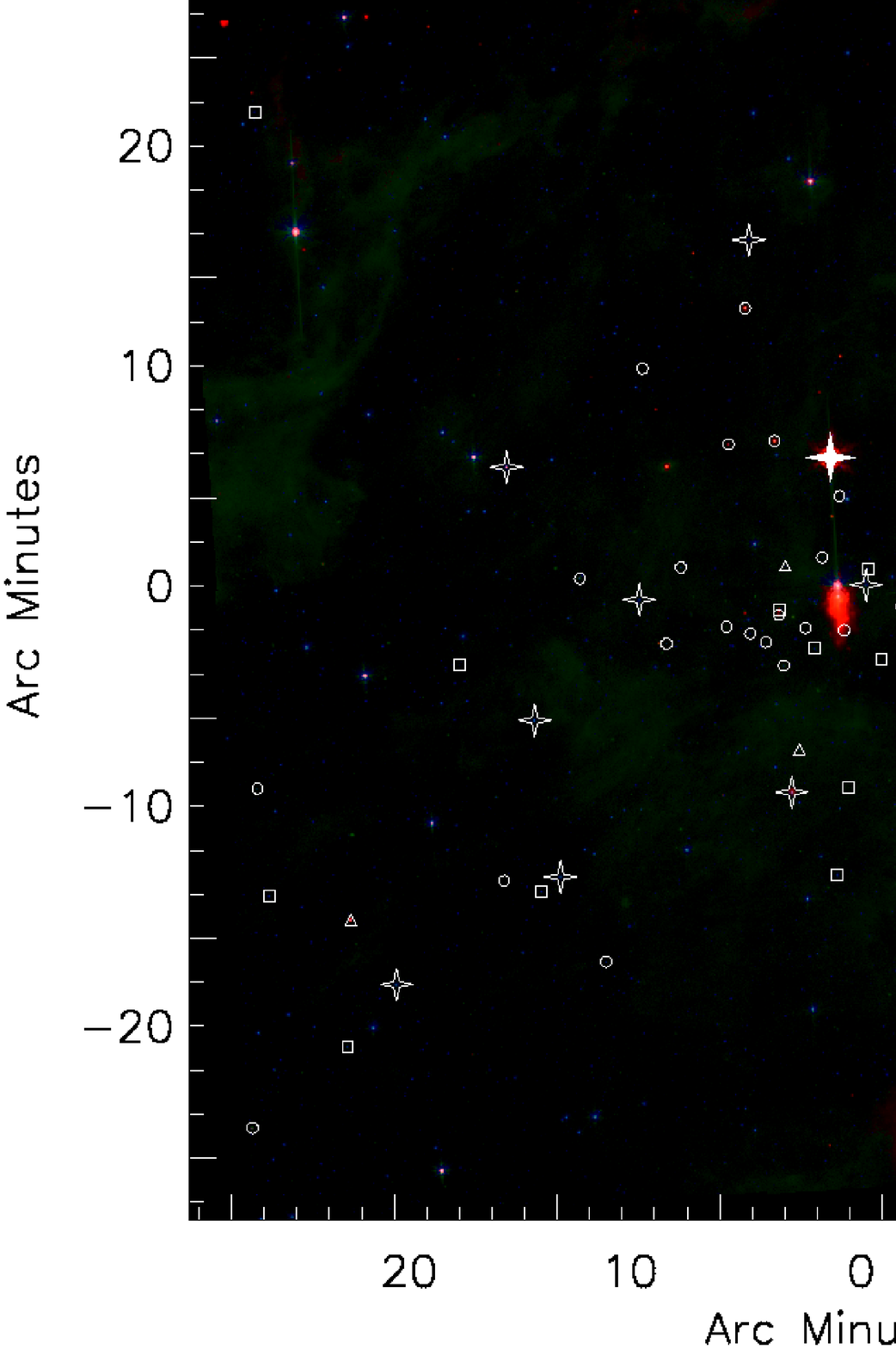}
\caption{The left panel shows the map of dust infrared emission 
\citep{schlegel98} of the {\LOri} star forming region,
with superimposed CO isocontours \citep{dame01}. 
Using spectral types in \citet{kharchenko09}, 
high mass stars with spectral type earlier than B5 are represented 
as large stars, while B-type stars with spectral type B5 or later 
are represented as small stars. The central box shows approximately
the IRAC field studied in this work. The right panel is a false color 
image of the {\LOri} cluster. It is a three-color composite of IRAC
images, 3.6 {\micron}(blue) and 8.0 \micron(green), and 
MIPS image, 24 \micron(red). The plot is centered at {\LOri}.
We show the location of stars bearing disks. Different symbols 
represent different types of disks classified in \S\ref{sec:res}: circles, squares and triangles represent
thick disks, evolved disks, and pre-transitional and transitional disks, respectively. 
Intermediate mass candidates surrounded by debris disks selected in \citetalias{hernandez09} 
are represented as open four-points stars. The solid four-point star indicates the Herbig Ae star HD 245185.}
\label{f:map}
\end{figure}

\begin{figure}
\epsscale{1.0}
\plotone{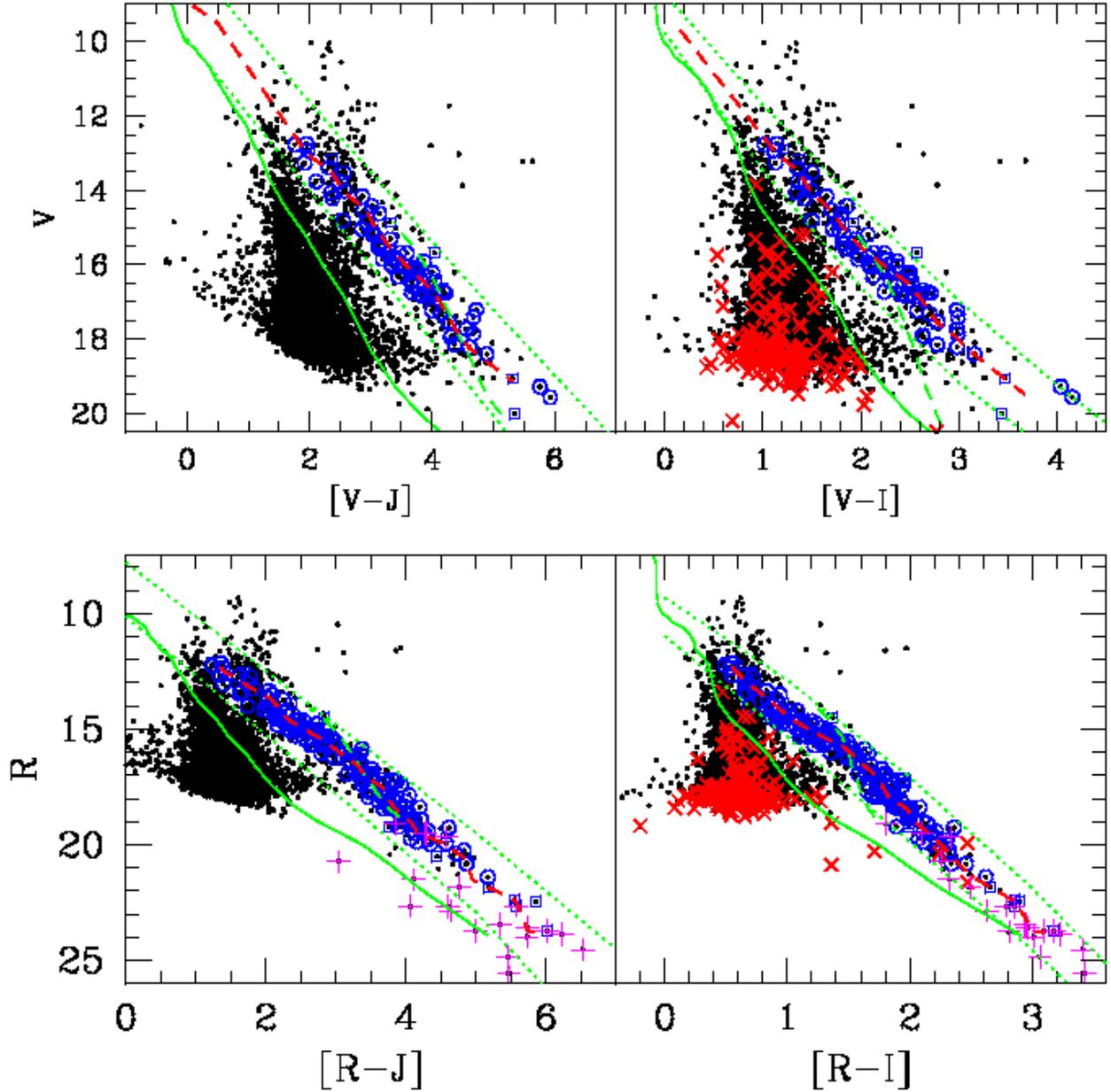}
\caption{Color-magnitude diagrams illustrating the selection of stars in the {\LOri} cluster 
($\lambda$Ori sample = known members + new photometric candidates). 
Open circles represent members confirmed using the presence of \ion{Li}{1} or radial velocity analysis. 
Open squares represent other members selected using spectroscopic data (see Table \ref{tab:known_mem}). 
Sources without NIR counterpart are represented with 
``x'' symbols. Sources with NIR photometry from \citetalias{barrado07b} are 
represented by crosses. Dashed lines represent the median of colors of the known members
in bins of 1 magnitude in the V-band (upper panels) 
or in the R-band (lower panels). Dotted lines 
limit the region
where members are expected to 
fall (member regions).  These member regions were created using distributions of the 
known members in each diagram. Assuming a distance of 450 pc, the ZAMS (solid line)
was plotted in each panel.} 
\label{f:cmd_sel}
\end{figure}

\begin{figure}
\epsscale{0.8}
\plotone{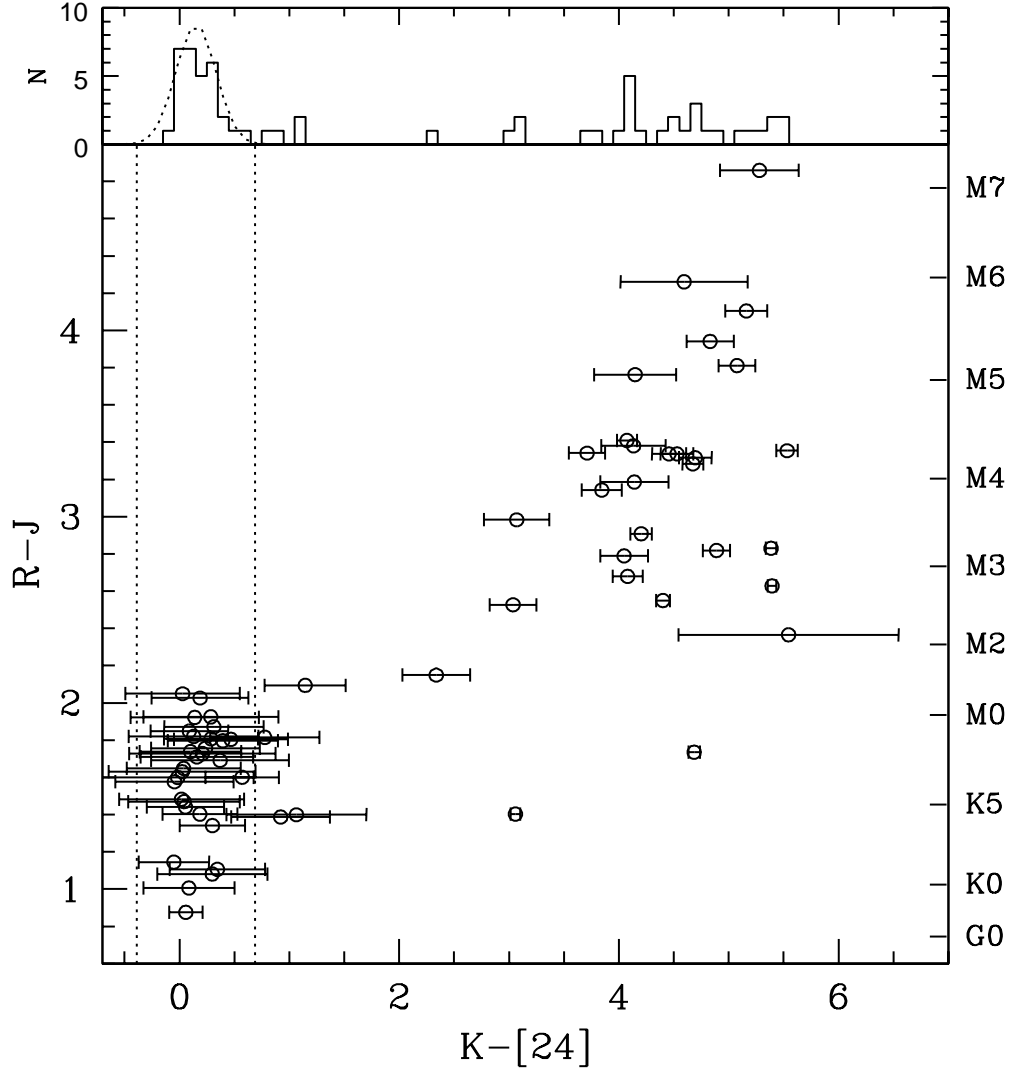}
\caption{ Color magnitude diagram of K-[24] versus R-J for the $\lambda$Ori sample. 
On the right vertical axis we are showing the spectral types that correspond 
to the standard colors from \citet[stars M5 or earlier;][]{kh95} and to the typical 
colors of members from \citetalias{barrado07b} (stars M6 or later). 
 The upper panel shows the distribution of K-[24] for stars 
in the $\lambda$Ori sample (\S\ref{s:photsel}). Most 
stars are located at K-[24]$\sim$0.15 exhibiting 
approximately a Gaussian distribution (dotted line) with $\sigma$ of 0.18 mag.
The 3$\sigma$ boundaries (dotted lines in lower panels) represent the photospheric region.}

\label{f:mips_excess}
\end{figure}

\begin{figure}
\epsscale{0.8}
\plotone{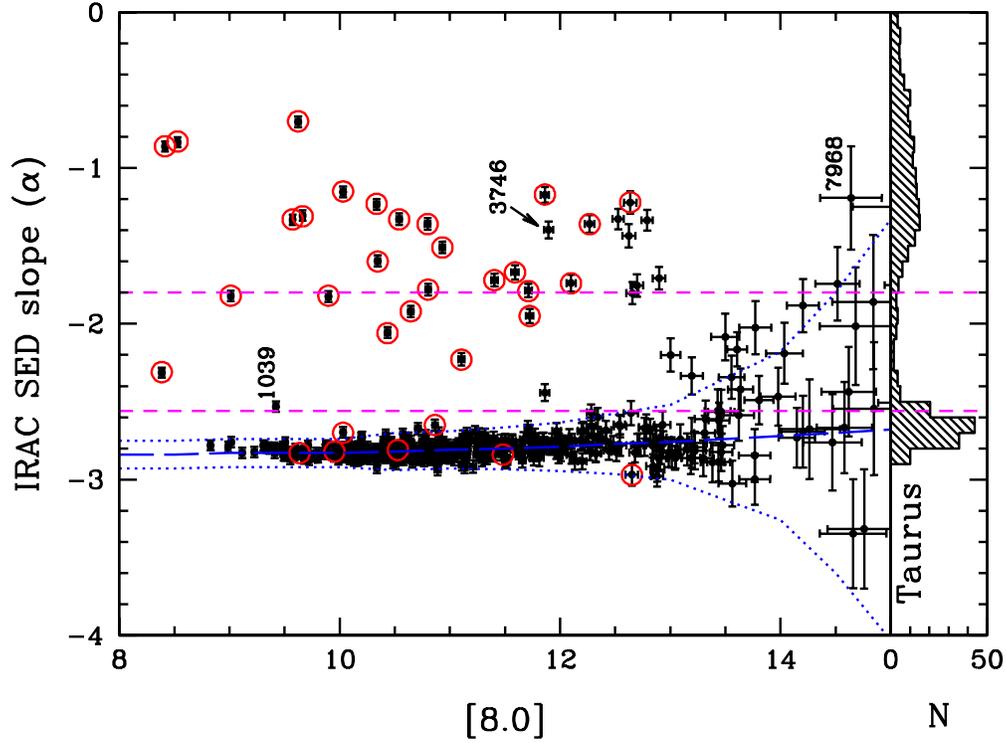}
\caption{IRAC SED slope versus [8.0] for the $\lambda$Ori sample. 
The slopes were calculated using 
the [3.6]-[8.0] color. Dotted lines represent the boundaries we have selected 
for the photospheric colors.
The lower dashed line represents the photometric 
limit used by other authors \citep[e.g.,][BN07]{lada06} to 
separate stars bearing disks 
and diskless stars. This limit is not appropriate for the faintest stars in our sample
because does not take into account photometric errors. 
The upper dashed line represents the limit between thick-disks and evolved disks objects 
from \citet{lada06}. 
The right-panel shows the IRAC SED slope distribution of the stellar population in Taurus \citep{luhman10}. 
Stars with 24{\micron} excess (see Figure \ref{f:mips_excess}) are represented by points surrounded by large
open circles.}
\label{f:irac}
\end{figure}

\begin{figure}
\epsscale{0.8}
\plotone{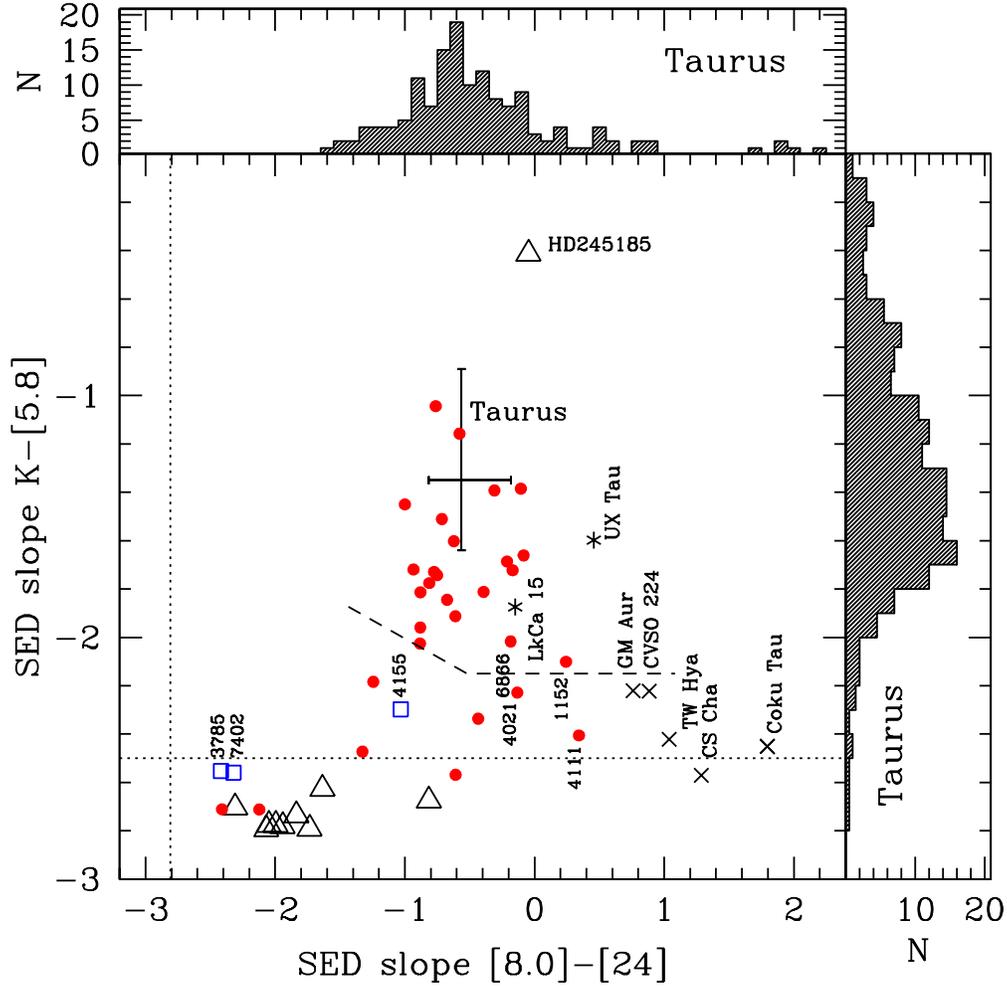}
\caption{SED slopes from  K-[5.8] and [8.0]-[24]
for disk bearing stars of the {\LOri} cluster with photometric 
spectral types of B-F \citepalias[triangles; ][]{hernandez09},
G-K4 (squares), and K5 or later (circles). As reference we display 
the location of 5 transitional disk objects (GM Aur, CVSO 224, TW Hya, CS Cha, Coku Tau/4) 
and 2 pre-transitional disk objects (LCa 15 and UX Tau A). 
Photospheric limits are indicated with dotted lines. 
Dashed line represents the lower boundary of primordial disks from \citet{luhman10}. 
This line is used as reference to separate stars bearing optically thick disks  
from other types of disk: transitional disk candidates ($\sim$ below the limit with SED slope [8.0]-[24]$>$0), 
pre-transitional disk candidates ($\sim$below the limit with 
SED slope [8.0]-[24]$\sim$0) and evolved disk systems ($\sim$below the limit with 
[8.0]-[24]$<$0). Early type stars below the photospheric limit of K-[5.8] are likely debris disk systems \citepalias[triangles; ][]{hernandez09}.
The right panel and upper panel show the SED slope distributions from K-[5.8] and [8.0]-[24] 
for the Taurus disks \citep{luhman10}; the error bar represents the median and quartiles for the Taurus disks }
\label{f:diskclass}
\end{figure}

\begin{figure}
\epsscale{0.8}
\plotone{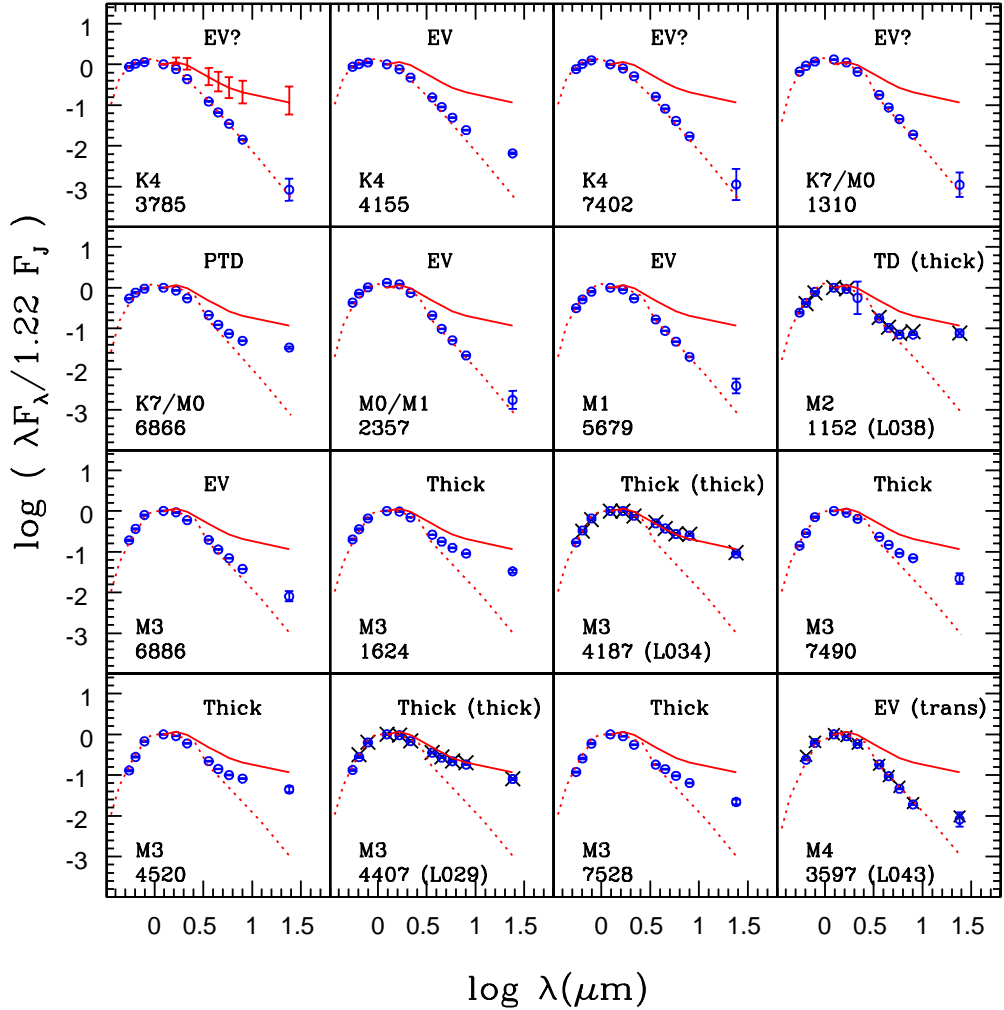}
\caption{SEDs for stars in the $\lambda$Ori sample with IR excesses and MIPS detections. 
In each panel, we show our identification number, the photometric spectral type and
the disk type found in \S \ref{disk_type}. Photometric spectral types are 
given using the R-J colors and standard calibrations.  Dotted lines show the 
corresponding photospheric levels \citep{kh95}.
For comparison, we show in parenthesis the identification number, the 
spectral type (when available) and the disk type for stars studied by
\citetalias{barrado07b}. The photometry from these authors are represented by crosses.
We plot the median SED slope of the disk population in Taurus \citep[solid line; ][]{luhman10};
error bars shown in the first panel denote the corresponding quartiles.}
\label{f:Ltype1}
\end{figure}

\begin{figure}
\epsscale{0.8}
\figurenum{6}
\plotone{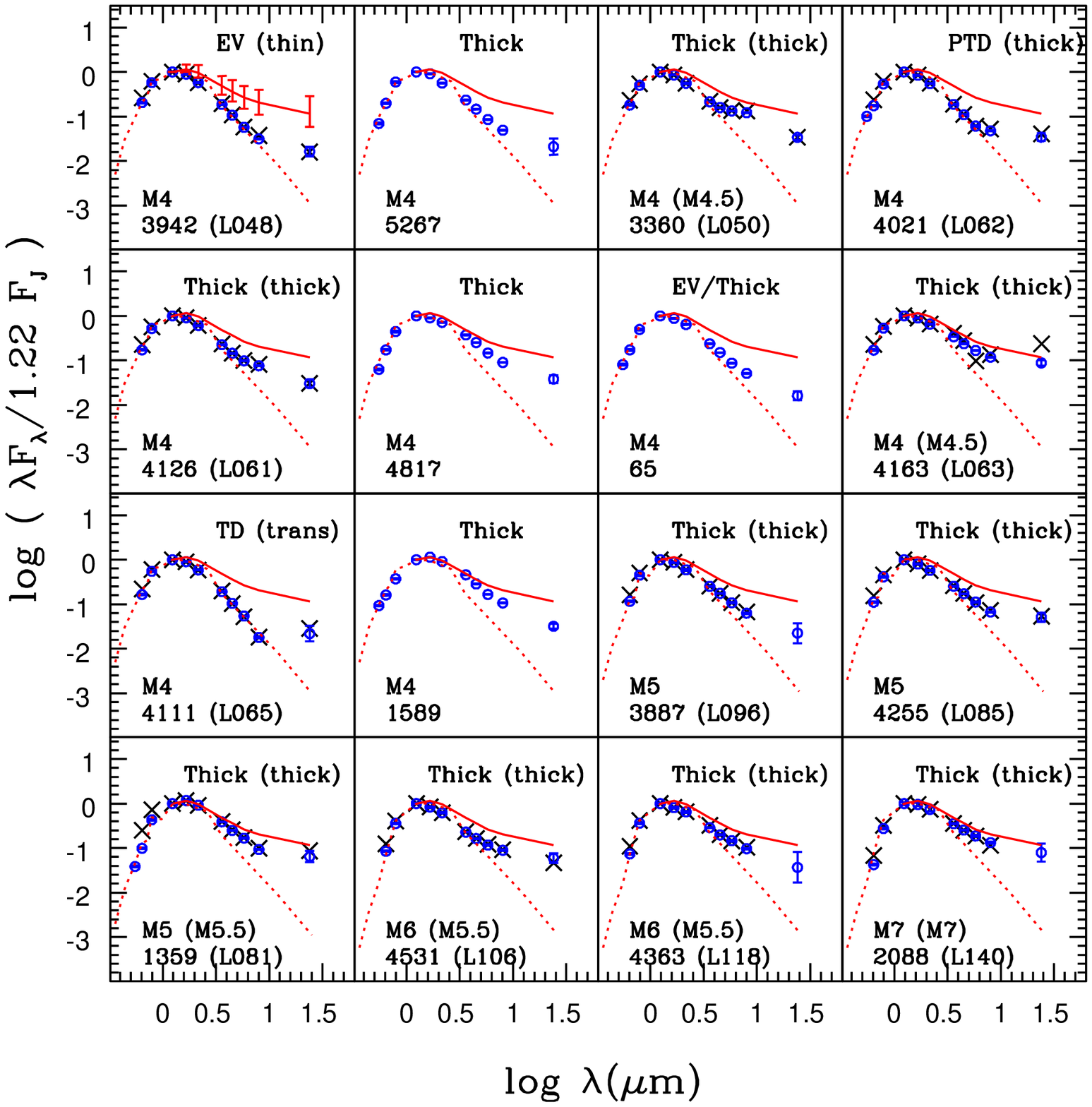}
\caption{SEDs for late type stars with IR excesses and MIPS detections.}
\end{figure}

\begin{figure}
\epsscale{0.8}
\plotone{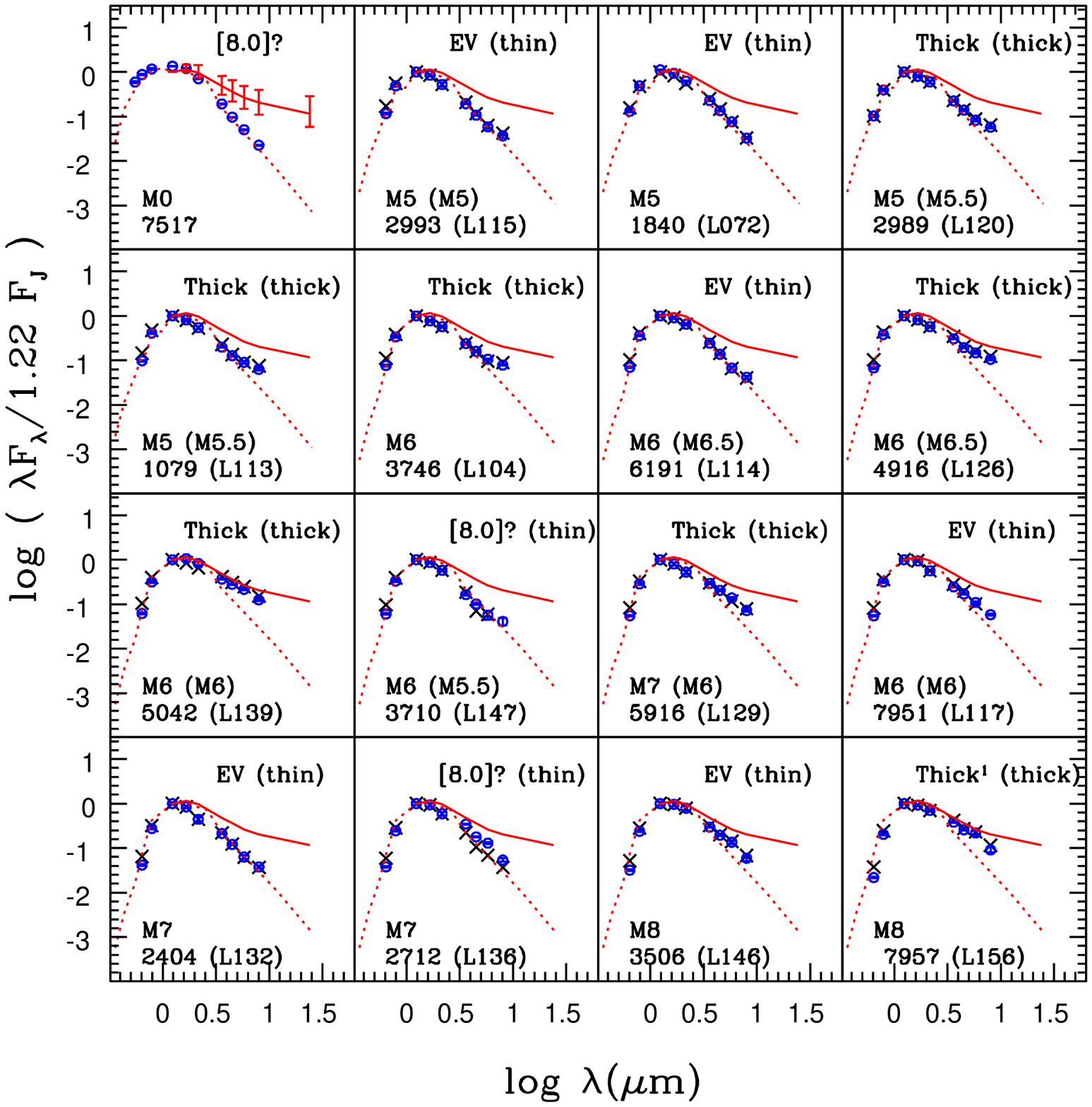}
\caption{SEDs for late type stars with IR excesses and without MIPS 
detections. Symbols and labels are similar to Figure \ref{f:Ltype1}.
From the IRAC SED slope, star \#7957 was classified bearing a thin disk.
The SED of this object suggests that \#7957 has an optically thick disks.}
\label{f:Ltype2}
\end{figure}

\begin{figure}
\epsscale{0.8}
\plotone{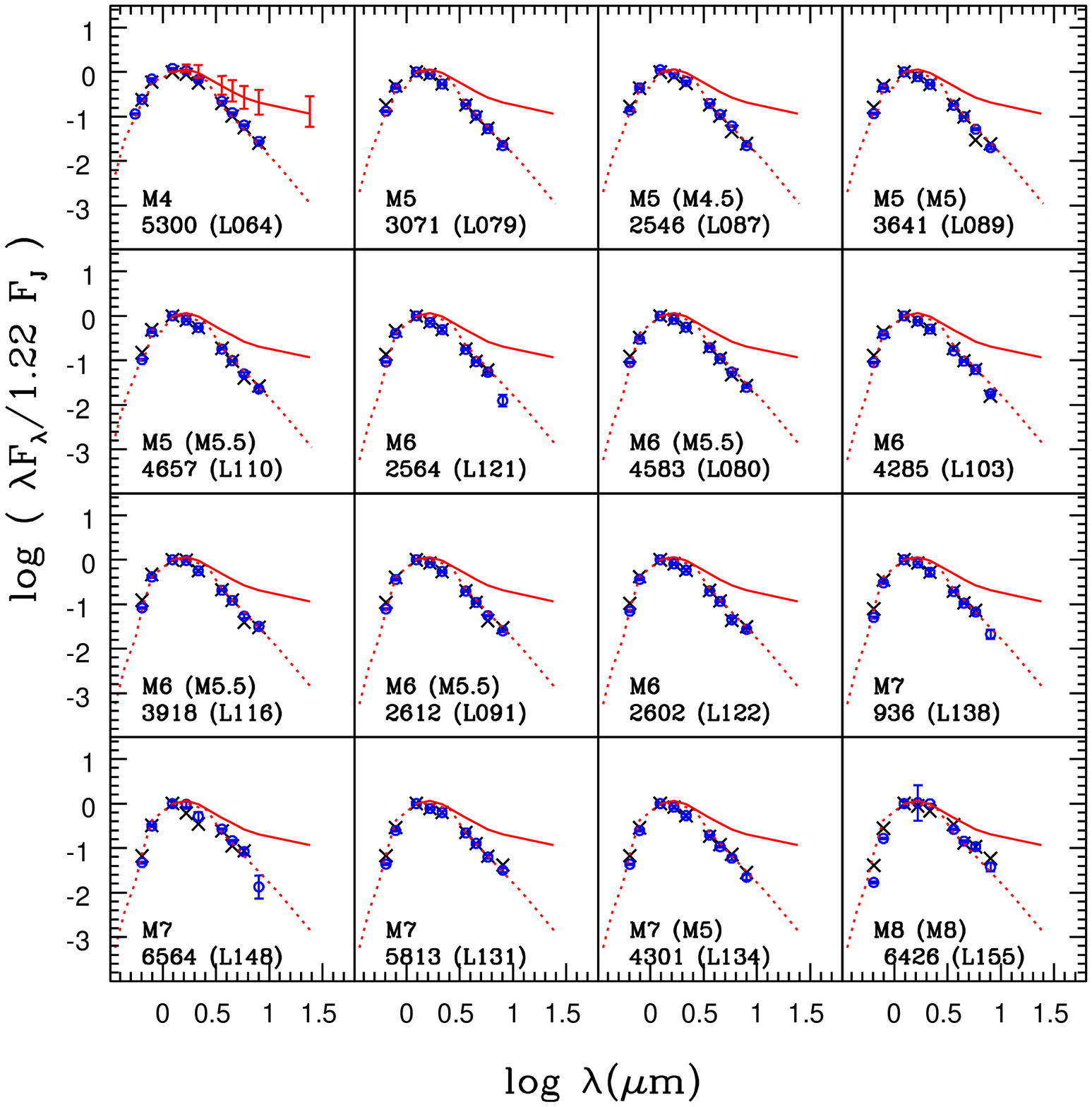}
\caption{SEDs for diskless stars classified previously as stars bearing 
thin disks \citepalias{barrado07b}. 
Symbols and labels are similar to Figure \ref{f:Ltype1}.}
\label{f:Ltype3}
\end{figure}

\begin{figure}
\epsscale{0.8}
\plotone{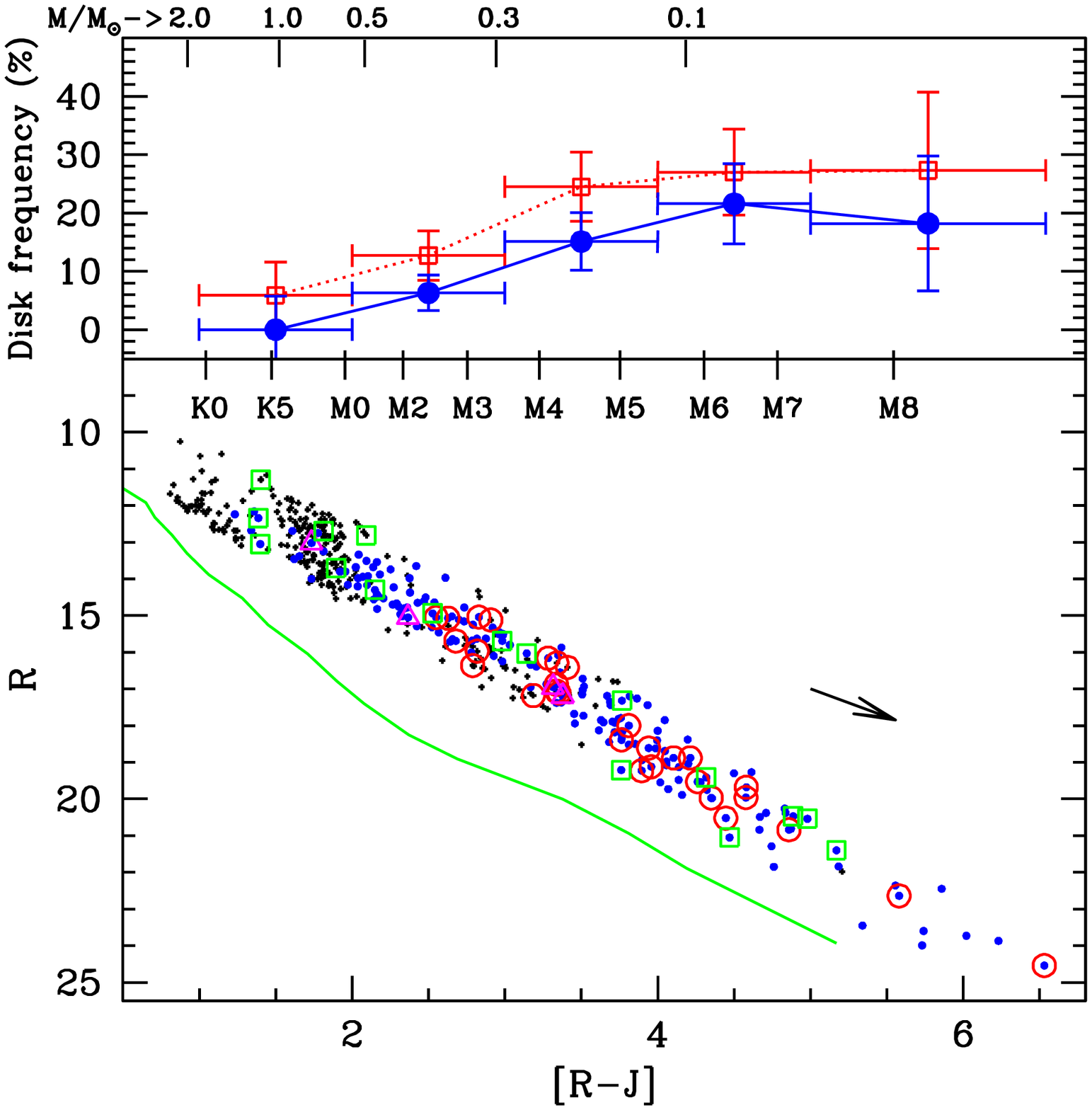}
\caption{
Upper panel:
Disk frequency as a function of photometric spectral type,
estimated using the \Rc-J color. The spectral type scale was calculated using the standard
colors from \citet{kh95} and the typical \Rc-J color of members from \citetalias{barrado07b}
with spectral type M6 or later (see \ref{mips_exc}). We display stellar masses using the
5 Myr isochrone from \citet{sf00}.  The disk frequencies for stars bearing optically thick 
disks (filled circles) and for all disk-bearing stars (open squares) decrease toward higher stellar masses.
Lower panel shows the locations of disk bearing stars of the {\LOri} cluster on a
color magnitude diagram.  Members from \citet{dolan01}, \citetalias{barrado07b}, \citet{sacco08} and \citet{maxted08}
are represented by solid circles; photometric candidates selected in \S\ref{s:photsel}
are represented by small crosses. 
Different open symbols represent different types of disks found in \S \ref{disk_type} 
and \S\ref{seds}; circles, squares and triangles represent {\it thick disks}, {\it evolved disks}
and {\it pre-transitional} and {\it transitional disks}, respectively.  
The solid line is the ZAMS \citep{sf00} located at 450 pc.  
}
\label{f:disk_frec}
\end{figure}

\begin{figure}
\epsscale{0.8}
\plotone{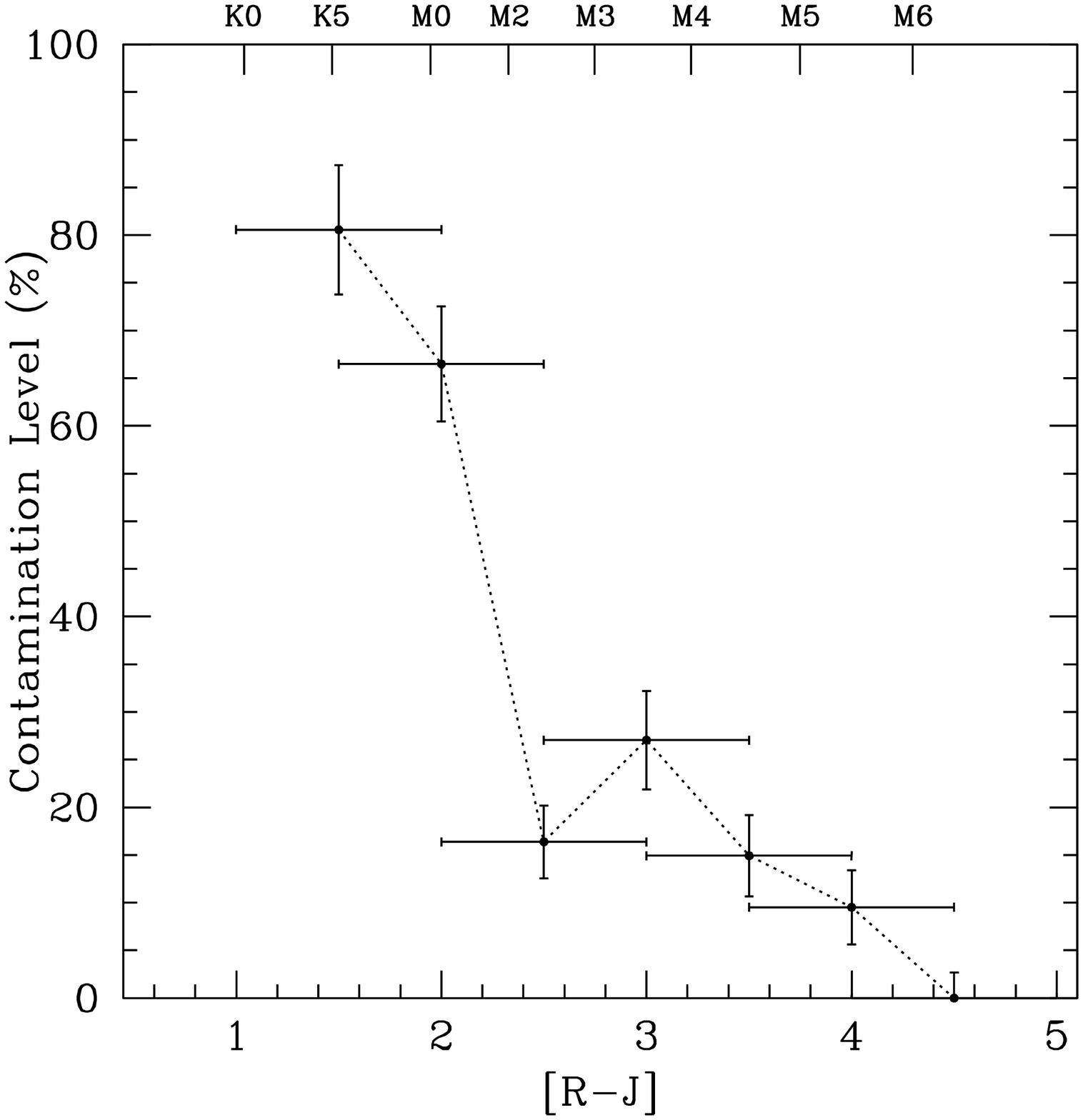}
\caption{Contamination level as a function of \Rc-J. 
To estimate the level of contamination by non-members in the $\lambda$Ori sample (\S\ref{s:photsel}), 
we calculate disk frequencies in this sample in several bins of colors ($F^{phot}_{disk}$). 
The disk frequencies assumed for the {\LOri} cluster ($F^{mem}_{disk}$) are calculated 
using the confirmed members from \citet{dolan01}, \citetalias{barrado07b}, \citet{sacco08}  
and \citet{maxted08}. The contamination level is 100*(1-$F^{phot}_{disk}$/$F^{mem}_{disk}$).}
\label{f:contamina}
\end{figure}

\begin{figure}
\epsscale{0.8}
\plotone{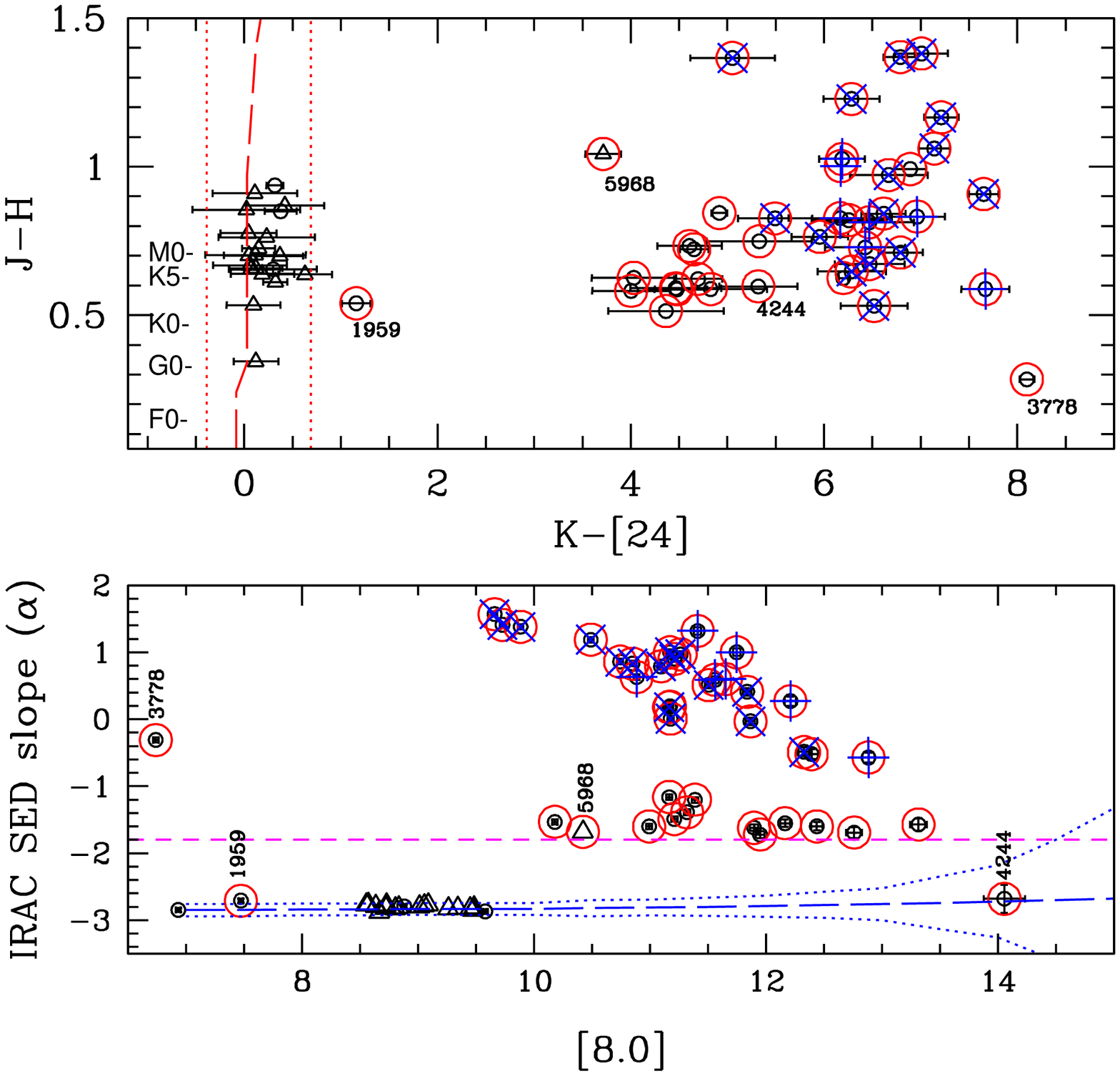}
\caption{Diagrams used to identify objects with excesses at
24{\micron} and 8{\micron} for MIPS sources not included in
the $\lambda$Ori sample. We plot 2MASS sources without 
optical photometric(circles) and sources with uncertain 
photometric membership (triangles).
The upper panel shows the color magnitude diagram K-[24] versus J-H (see \S\ref{mips_exc}).
We display the limits used in Figure \ref{f:mips_excess}. 
The lower panel shows the IRAC SED slope versus [8.0] diagram K-[24] (see \S\ref{irac_exc}).
We display the limits used in Figure \ref{f:irac}.
Objects surrounded by large circles represent sources with excesses at 24{\micron. 
We show possible galaxies (symbol X) and other diffuse object (crosses) 
selected by visual inspection.  }.
}
\label{f:disk_add}
\end{figure}

\end{document}

%% file: tab1s.tex
\begin{deluxetable}{cccccccccccc}
\rotate
\tabletypesize{\scriptsize}
\tablewidth{0pt}
\tablecaption{Compiled membership information of the $\lambda$ Orionis Cluster \label{tab:known_mem}}
\tablehead{
\colhead{ID} & \colhead{RA(2000)} & \colhead{DEC(2000)} & \colhead{name(s)} & \colhead{Spectral} & \colhead{Ref.} & \colhead{radial vel.} & \colhead{Ref.} & \colhead{${\rm Li_{~I}}$} & \colhead{Ref.}     & \colhead{membership} & \colhead{membership} \\
\colhead{ } & \colhead{deg}      & \colhead{deg}        & \colhead{ }       & \colhead{Type}     & \colhead{(SpT)}& \colhead{km s$^{-1}$}  & \colhead{r.v.} & \colhead{present}    & \colhead{${\rm Li_{~I}}$} &\colhead{reference} & \colhead{flag} \\
\colhead{(1)} & \colhead{(2)}      & \colhead{(3)}     & \colhead{(4)}       & \colhead{(5)}     & \colhead{(6)}& \colhead{(7)}  & \colhead{(8)} & \colhead{(9)}    & \colhead{(10)} &\colhead{(11)} & \colhead{(12)} \\
}
\startdata
 936 & 83.43113 &  9.75636 &         LOri138 &          \nodata &      \nodata &          \nodata &      \nodata &    \nodata &   \nodata &         2a &    m5 \\
1048 & 83.44671 &  9.92733 &     DM1,LOri001 &          \nodata &      \nodata &        22.44 &        1 &      Y &     1 &      1a,2a &    m1 ,\\
1079 & 83.44962 & 10.02767 &         LOri113 &         M5.5 &        2 &          \nodata &      \nodata &    \nodata &   \nodata &        2a &    m4 \\
1152 & 83.45800 &  9.84353 &     DM2,LOri038 &          \nodata &      \nodata &        24.03 &        1 &      Y &     1 &      1a,2a &    m1 \\
1213 & 83.46550 &  9.63928 &             DM3 &          \nodata &      \nodata &        21.27 &        1 &      Y &     1 &         1a &    m1 \\
1276 & 83.47354 &  9.71900 &         LOri005 &          \nodata &      \nodata &          \nodata &      \nodata &    \nodata &   \nodata &         2a &    m5 \\
1346 & 83.48471 &  9.89911 &     DM4,LOri013 &          \nodata &      \nodata &         2.40 &        1 &      Y &     1 &      1b,2a &    m3 \\
1359 & 83.48587 & 10.10414 &         LOri081 &         M5.5 &        2 &          \nodata &      \nodata &    \nodata &   \nodata &        2a &    m4 \\
1531 & 83.50838 &  9.68503 &             DM5 &          \nodata &      \nodata &        28.98 &        1 &      Y &     1 &         1a &    m1 \\
1728 & 83.53496 &  9.85703 &         LOri044 &          \nodata &      \nodata &         6.52 &        3 &    \nodata &   \nodata &       2a,3b &   nm2 \\
1733 & 83.53554 &  9.84542 &         LOri109 &         M5.5 &        2 &        28.10 &        3 &    \nodata &   \nodata &      2a,3a &    m2 \\
1829 & 83.54658 &  9.85833 &         LOri127 &          \nodata &      \nodata &          \nodata &      \nodata &    \nodata &   \nodata &         2d &   nm1 \\
1835 & 83.54704 &  9.74078 &         LOri152 &          \nodata &      \nodata &          \nodata &      \nodata &    \nodata &   \nodata &         2c &   nm1 \\
1840 & 83.54754 &  9.70219 &         LOri072 &          \nodata &      \nodata &          \nodata &      \nodata &    \nodata &   \nodata &         2a &    m5 \\
1844 & 83.54825 &  9.82083 &         LOri086 &          \nodata &      \nodata &        30.01 &        3 &    \nodata &   \nodata &       2a,3a &    m2 \\
1847 & 83.54904 &  9.95092 &         LOri052 &          \nodata &      \nodata &        36.31 &        3 &    \nodata &   \nodata &       2a,3b &   nm2 \\
1926 & 83.55933 &  9.80731 &     LOri124-125 &         M5.5 &        2 &        30.76 &        3 &    \nodata &   \nodata &    2a,3a &    m2 \\
7971 & 83.90367 &  9.74011 &         LOri170 &          \nodata &      \nodata &          \nodata &      \nodata &    \nodata &   \nodata &         2c &   nm1 \\
\enddata
\tablecomments{Table \ref{tab:known_mem} is published in its entirety in the electronic edition of
the {\it Astrophysical Journal}. A portion is shown here for guidance
regarding its form and content.}
\tablenotetext{~}{References: (1) \citet{dolan01}; (2) \citet{barrado07b}; (3) \citet{maxted08}; (4) \citet{sacco08} }
\tablenotetext{1}{Column 11: (1a) and (1b) represent stars with ${\rm Li_{~I}}$ in absorption from \citet{dolan01} and Radial Velocity in the members range and  out the member range,respectively;
(2a), (2b), (2c) and (2d) represent the member flags Y, Y?, N?, and N, respectively, from \citet{barrado07b}; (3a) members, (3b) no members from the radial velocity analysis of \citet{maxted08}; (4a) and (4b) member and no members from 
the ${\rm Li_{~I}}$ and radial velocity analysis of \citet{sacco08}}
\tablenotetext{2}{Column 12: (m1) radial velocity members with ${\rm Li_{~I}}$ in absorption; (m2) Radial velocity members with no ${\rm Li_{~I}}$ information; (m3) stars with ${\rm Li_{~I}}$ in absorption and radial velocity out of member range; 
(m4) Both spectral type and photometric data in aggrement with cluster 
sequence; (m5) photometric members; (nm1) non-members based on photometric 
analysis; (nm2) non-members based in spectroscopic analysis }
\end{deluxetable}

%% file: tab2s.tex
\begin{deluxetable}{ccccccccccc}
\rotate
\tabletypesize{\scriptsize}
\tablewidth{700pt}
\tablecaption{Photometric candidates of the $\lambda$ Orionis Cluster \label{t:main}}
\tablehead{
\colhead{ID} &\colhead{2MASS} &\colhead{RA(2000)} & \colhead{DEC(2000)} & \colhead{[3.6] } & \colhead{[4.5]} & \colhead{[5.8]} & \colhead{[8.0]} & \colhead{Flux$_{24\micron}$} & \colhead{Disk type} & \colhead{membership}  \\
\colhead{ } & \colhead{ } & \colhead{deg} & \colhead{deg} & \colhead{mag} & \colhead{mag} & \colhead{mag}& \colhead{mag}  & \colhead{mJy} & \colhead{}    & \colhead{reference\tablenotemark{8}} \\
\colhead{(1)} & \colhead{(2)} & \colhead{(3)} & \colhead{(4)} & \colhead{(5)} & \colhead{(6)} & \colhead{(7)}& \colhead{(8)}  & \colhead{(9)} & \colhead{(10)}    & \colhead{(11)} \\
}
\startdata 
  30 & 05331224+0934464 & 83.30102 &  9.57958 &  10.71 $\pm$  0.02 &  10.71 $\pm$  0.02 &  10.68 $\pm$  0.02 &  10.65 $\pm$  0.02 &   -99.99 $\pm$ 99.990 & diskless & \nodata \\  
  42 & 05331308+0937597 & 83.30452 &  9.63328 &  11.50 $\pm$  0.02 &  11.52 $\pm$  0.02 &  11.43 $\pm$  0.03 &  11.45 $\pm$  0.03 &   -99.99 $\pm$ 99.990 & diskless & \nodata \\  
  65 & 05331515+0950301 & 83.31313 &  9.84172 &  11.45 $\pm$  0.02 &  11.24 $\pm$  0.02 &  11.08 $\pm$  0.03 &  10.64 $\pm$  0.02 &     3.96 $\pm$  0.890 & Ev/Thk & \nodata \\  
 208 & 05332084+0958048 & 83.33684 &  9.96803 &  11.69 $\pm$  0.02 &  11.73 $\pm$  0.02 &  11.63 $\pm$  0.03 &  11.63 $\pm$  0.03 &   -99.99 $\pm$ 99.990 & diskless & \nodata \\  
 296 & 05332402+1019535 & 83.35012 & 10.33153 &  10.78 $\pm$  0.02 &  10.75 $\pm$  0.02 &  10.81 $\pm$  0.02 &  10.77 $\pm$  0.02 &   -99.99 $\pm$ 99.990 & diskless & \nodata \\  
 328 & 05332480+1012116 & 83.35334 & 10.20324 &  10.30 $\pm$  0.02 &  10.26 $\pm$  0.02 &  10.22 $\pm$  0.02 &  10.20 $\pm$  0.02 &   -99.99 $\pm$ 99.990 & diskless & \nodata \\  
 623 & 05333353+0935494 & 83.38974 &  9.59708 &  10.44 $\pm$  0.02 &  10.43 $\pm$  0.02 &  10.46 $\pm$  0.02 &  10.42 $\pm$  0.02 &   -99.99 $\pm$ 99.990 & diskless & \nodata \\  
 630 & 05333378+1009233 & 83.39077 & 10.15650 &  10.73 $\pm$  0.02 &  10.78 $\pm$  0.02 &  10.76 $\pm$  0.02 &  10.62 $\pm$  0.02 &   -99.99 $\pm$ 99.990 & diskless & \nodata \\  
7662 & 05370084+0949045 & 84.25352 &  9.81792 &   9.94 $\pm$  0.02 &   9.93 $\pm$  0.02 &   9.87 $\pm$  0.02 &   9.88 $\pm$  0.02 &   -99.99 $\pm$ 99.990 & diskless & \nodata \\  
7689 & 05370204+1014447 & 84.25851 & 10.24577 &  11.07 $\pm$  0.02 &  11.12 $\pm$  0.02 &  11.10 $\pm$  0.03 &  11.11 $\pm$  0.03 &   -99.99 $\pm$ 99.990 & diskless & \nodata \\  
7701 & 05370265+1003583 & 84.26104 & 10.06620 &  11.49 $\pm$  0.02 &  11.52 $\pm$  0.02 &  11.47 $\pm$  0.03 &  11.52 $\pm$  0.03 &   -99.99 $\pm$ 99.990 & diskless & \nodata \\  
7957 & \nodata & 83.65117 &  9.92561 &  15.04 $\pm$  0.03 &  14.76 $\pm$  0.04 &  14.19 $\pm$  0.13 &  14.20 $\pm$  0.14 &   -99.99 $\pm$ 99.990 & Thk\tablenotemark{3,6} & 2a \\  
7968 & \nodata & 83.80912 &  9.90209 &  16.08 $\pm$  0.07 &  15.86 $\pm$  0.08 &  16.23 $\pm$  0.73 &  14.64 $\pm$  0.28 &   -99.99 $\pm$ 99.990 & diskless\tablenotemark{4} & 2a \\  
\enddata
\tablecomments{Table \ref{t:main} is published in its entirety in the electronic edition of
the {\it Astrophysical Journal}. A portion is shown here for guidance regarding its form and content.}
\tablecomments{Column 10: Thk-star with an optically thick disk; TD-transitional disk candidate; 
PTD-pre transitional disk candidate; EV-star with an evolved (flat) disk; diskless-star with no infrared excesses.}
\tablenotetext{1}{stars with uncertain 24{\micron} excess}
\tablenotetext{2}{stars with uncertain 8{\micron} excess}
\tablenotetext{3}{stars without 24{\micron} counterpart}
\tablenotetext{4}{Faint stars with not reliable excesses at 5.8{\micron} \& 8.0{\micron}}
\tablenotetext{5}{Star with moderate IR excess at 8.0{\micron} but the 24{\micron} flux 
is consistent with the photospheric fluxes}
\tablenotetext{6}{Classified as evolved disk star from the IRAC SED slope}
\tablenotetext{7}{The star {\LOri} may mask the detection at 24\micron}
\tablenotetext{8}{See Table \ref{tab:known_mem}}
\end{deluxetable}

%% file: tab3.tex
\begin{deluxetable}{cccccccccc}
\rotate
\tabletypesize{\scriptsize}
\tablewidth{0pt}
\tablecaption{Additional sources with infrared excesses \label{t:append}}
\tablehead{
\colhead{ID} &\colhead{2MASS} &\colhead{RA(2000)} & \colhead{DEC(2000)} & \colhead{[3.6] } & \colhead{[4.5]} & \colhead{[5.8]} & \colhead{[8.0]} & \colhead{Flux$_{24\micron}$} & \colhead{Comments}  \\
\colhead{ } & \colhead{ } & \colhead{deg} & \colhead{deg} & \colhead{mag} & \colhead{mag} & \colhead{mag}& \colhead{mag}  & \colhead{mJy} & \colhead{} \\
\colhead{(1)} & \colhead{(2)} & \colhead{(3)} & \colhead{(4)} & \colhead{(5)} & \colhead{(6)} & \colhead{(7)}& \colhead{(8)}  & \colhead{(9)} & \colhead{(10)} \\
}
\startdata 
241 & 05332198+1002019 & 83.34160 & 10.03387 & 12.928 $\pm$ 0.022 & 12.700 $\pm$ 0.023 & 12.464 $\pm$ 0.038 & 11.949 $\pm$ 0.038 & 1.42 $\pm$ 0.85 &	 \\
336 & 05332499+0937360 & 83.35415 & 9.62667 & 13.658 $\pm$ 0.025 & 13.439 $\pm$ 0.028 & 13.467 $\pm$ 0.082 & 11.177 $\pm$ 0.027 & 1.89 $\pm$ 0.98 & Galaxy \\
387 & 05332696+1001394 & 83.36234 & 10.02764 & 14.417 $\pm$ 0.031 & 13.833 $\pm$ 0.031 & 13.306 $\pm$ 0.068 & 12.393 $\pm$ 0.054 & 1.87 $\pm$ 0.93 &	 \\
415 & 05332785+0933527 & 83.36605 & 9.56464 & 14.469 $\pm$ 0.030 & 14.366 $\pm$ 0.037 & 13.560 $\pm$ 0.070 & 11.211 $\pm$ 0.027 & 2.00 $\pm$ 0.86 & Galaxy\\
547 & 05333119+0958575 & 83.38000 & 9.98265 & 14.919 $\pm$ 0.034 & 14.793 $\pm$ 0.042 & 14.491 $\pm$ 0.168 & 12.210 $\pm$ 0.049 & 1.79 $\pm$ 0.95 & Stellar? \\
685 & 05333531+1008261 & 83.39715 & 10.14059 & 14.425 $\pm$ 0.029 & 14.208 $\pm$ 0.034 & 13.938 $\pm$ 0.084 & 13.314 $\pm$ 0.073 & 1.11 $\pm$ 0.82 &	 \\
907 & 05334256+1003478 & 83.42737 & 10.06328 & 13.812 $\pm$ 0.025 & 13.169 $\pm$ 0.025 & 12.567 $\pm$ 0.040 & 11.172 $\pm$ 0.026 & 5.16 $\pm$ 0.85 &	 \\
1110 & 05334885+0959544 & 83.45355 & 9.99847 & 14.251 $\pm$ 0.029 & 14.076 $\pm$ 0.033 & 13.710 $\pm$ 0.079 & 11.094 $\pm$ 0.030 & 2.12 $\pm$ 0.98 & Galaxy\\
1959 & 05341507+1019091 & 83.56283 & 10.31921 & 7.599 $\pm$ 0.020 & 7.492 $\pm$ 0.020 & 7.458 $\pm$ 0.020 & 7.472 $\pm$ 0.020 & 9.07 $\pm$ 1.60 &	 \\
2040 & 05341771+0939224 & 83.57379 & 9.65623 & 12.591 $\pm$ 0.022 & 12.295 $\pm$ 0.022 & 11.936 $\pm$ 0.031 & 11.319 $\pm$ 0.028 & 1.59 $\pm$ 0.88 &	 \\
2191 & 05342233+1023279 & 83.59305 & 10.39108 & 15.027 $\pm$ 0.036 & 14.506 $\pm$ 0.036 & 14.215 $\pm$ 0.114 & 11.410 $\pm$ 0.032 & 3.08 $\pm$ 0.89 & Stellar?\\
2773 & 05344013+0956435 & 83.66724 & 9.94544 & 13.494 $\pm$ 0.024 & 13.355 $\pm$ 0.027 & 12.472 $\pm$ 0.047 & 9.659 $\pm$ 0.022 & 11.80 $\pm$ 1.39 & Galaxy\\
3212 & 05345260+0955500 & 83.71918 & 9.93058 & 11.323 $\pm$ 0.021 & 10.967 $\pm$ 0.021 & 10.677 $\pm$ 0.023 & 10.179 $\pm$ 0.022 & 9.22 $\pm$ 1.02 &	 \\
3245 & 05345340+0948500 & 83.72252 & 9.81390 & 13.963 $\pm$ 0.026 & 13.794 $\pm$ 0.029 & 13.382 $\pm$ 0.071 & 10.742 $\pm$ 0.025 & 4.16 $\pm$ 1.07 & Galaxy \\
3268 & 05345389+0946268 & 83.72455 & 9.77411 & 14.546 $\pm$ 0.030 & 14.046 $\pm$ 0.032 & 13.646 $\pm$ 0.080 & 11.561 $\pm$ 0.032 & 5.27 $\pm$ 0.82 & Stellar? \\
3323 & 05345526+0935368 & 83.73026 & 9.59357 & 14.425 $\pm$ 0.030 & 14.036 $\pm$ 0.031 & 13.861 $\pm$ 0.098 & 11.505 $\pm$ 0.033 & 2.24 $\pm$ 0.97 & Galaxy \\
3451 & 05345860+0937287 & 83.74419 & 9.62465 & 14.668 $\pm$ 0.030 & 14.222 $\pm$ 0.032 & 13.861 $\pm$ 0.090 & 11.837 $\pm$ 0.037 & 2.06 $\pm$ 0.87 & Galaxy \\
3778 & 05350816+0955344 & 83.78401 & 9.92623 & 8.945 $\pm$ 0.020 & 8.799 $\pm$ 0.021 & 8.390 $\pm$ 0.024 & 6.738 $\pm$ 0.023 & 3149.89 $\pm$ 313.33 &	 \\
3786 & 05350835+0935535 & 83.78479 & 9.59822 & 12.968 $\pm$ 0.022 & 12.754 $\pm$ 0.023 & 12.597 $\pm$ 0.042 & 11.896 $\pm$ 0.035 & 1.96 $\pm$ 0.84 &	 \\
3822 & 05350931+0959112 & 83.78880 & 9.98646 & 13.424 $\pm$ 0.024 & 13.351 $\pm$ 0.028 & 12.466 $\pm$ 0.041 & 9.727 $\pm$ 0.022 & 8.22 $\pm$ 1.42 & Galaxy \\
3910 & 05351190+0951178 & 83.79960 & 9.85495 & 13.795 $\pm$ 0.028 & 13.756 $\pm$ 0.033 & 13.541 $\pm$ 0.081 & 11.164 $\pm$ 0.034 & 1.88 $\pm$ 0.64 & Galaxy \\
3933 & 05351228+0955131 & 83.80120 & 9.92033 & 12.079 $\pm$ 0.022 & 11.719 $\pm$ 0.022 & 11.429 $\pm$ 0.028 & 10.992 $\pm$ 0.026 & 3.17 $\pm$ 1.45 &	 \\
4013 & 05351511+0943596 & 83.81300 & 9.73324 & 14.648 $\pm$ 0.032 & 14.249 $\pm$ 0.036 & 13.982 $\pm$ 0.101 & 11.653 $\pm$ 0.034 & 2.43 $\pm$ 0.88 & Stellar?\\
4057 & 05351649+0954372 & 83.81873 & 9.91034 & 13.555 $\pm$ 0.025 & 13.356 $\pm$ 0.027 & 12.502 $\pm$ 0.039 & 9.883 $\pm$ 0.022 & 5.43 $\pm$ 1.04 & Galaxy \\
4244 & 05352114+1005584 & 83.83810 & 10.09958 & 14.206 $\pm$ 0.027 & 14.169 $\pm$ 0.033 & 14.065 $\pm$ 0.099 & 14.055 $\pm$ 0.179 & 1.48 $\pm$ 0.81 &	 \\
4321 & 05352343+0955335 & 83.84766 & 9.92599 & 14.512 $\pm$ 0.029 & 14.199 $\pm$ 0.033 & 13.823 $\pm$ 0.097 & 11.172 $\pm$ 0.028 & 2.49 $\pm$ 0.50 & Galaxy \\
4405 & 05352535+0954476 & 83.85565 & 9.91324 & 12.630 $\pm$ 0.022 & 12.329 $\pm$ 0.022 & 12.008 $\pm$ 0.032 & 11.165 $\pm$ 0.026 & 12.28 $\pm$ 0.59 &	 \\
4527 & 05352872+0934328 & 83.86968 & 9.57579 & 13.292 $\pm$ 0.023 & 13.098 $\pm$ 0.025 & 12.877 $\pm$ 0.048 & 12.164 $\pm$ 0.042 & 1.32 $\pm$ 0.85 &	 \\
4693 & 05353347+1002087 & 83.88948 & 10.03576 & 14.380 $\pm$ 0.032 & 13.985 $\pm$ 0.033 & 14.192 $\pm$ 0.118 & 12.327 $\pm$ 0.047 & 0.94 $\pm$ 0.54 & Galaxy\\
5045 & 05354444+0945520 & 83.93517 & 9.76446 & 12.392 $\pm$ 0.021 & 12.133 $\pm$ 0.022 & 11.913 $\pm$ 0.030 & 11.209 $\pm$ 0.026 & 4.40 $\pm$ 0.88 &	 \\
5367 & 05355400+1003458 & 83.97500 & 10.06274 & 14.576 $\pm$ 0.032 & 14.069 $\pm$ 0.033 & 14.078 $\pm$ 0.121 & 11.263 $\pm$ 0.026 & 3.37 $\pm$ 0.92 & Galaxy\\
5541 & 05355903+1013175 & 83.99598 & 10.22153 & 13.765 $\pm$ 0.024 & 13.527 $\pm$ 0.026 & 13.332 $\pm$ 0.064 & 12.759 $\pm$ 0.068 & 1.06 $\pm$ 0.84 &	 \\
5582 & 05360034+0956488 & 84.00143 & 9.94691 & 14.045 $\pm$ 0.026 & 13.960 $\pm$ 0.030 & 13.387 $\pm$ 0.069 & 10.849 $\pm$ 0.024 & 2.48 $\pm$ 0.87 & Galaxy\\
5968 & 05361125+0946263 & 84.04689 & 9.77397 & 11.437 $\pm$ 0.021 & 11.045 $\pm$ 0.021 & 10.733 $\pm$ 0.023 & 10.422 $\pm$ 0.023 & 3.35 $\pm$ 0.85 &	 \\
6047 & 05361360+0930386 & 84.05668 & 9.51073 & 14.862 $\pm$ 0.033 & 14.408 $\pm$ 0.034 & 14.287 $\pm$ 0.116 & 12.883 $\pm$ 0.054 & 1.53 $\pm$ 0.83 & Stellar?\\
6638 & 05363109+0953262 & 84.12955 & 9.89062 & 13.904 $\pm$ 0.027 & 13.739 $\pm$ 0.030 & 13.495 $\pm$ 0.068 & 10.884 $\pm$ 0.024 & 2.95 $\pm$ 0.87 & Stellar?\\
6797 & 05363604+0934545 & 84.15020 & 9.58182 & 13.527 $\pm$ 0.024 & 13.337 $\pm$ 0.026 & 13.110 $\pm$ 0.053 & 12.438 $\pm$ 0.048 & 1.01 $\pm$ 0.83 &	 \\
7039 & 05364332+0957016 & 84.18051 & 9.95045 & 13.992 $\pm$ 0.026 & 13.785 $\pm$ 0.030 & 13.343 $\pm$ 0.058 & 10.489 $\pm$ 0.023 & 6.50 $\pm$ 1.01 & Galaxy \\
7179 & 05364695+0933541 & 84.19565 & 9.56505 & 15.090 $\pm$ 0.034 & 14.993 $\pm$ 0.048 & 14.735 $\pm$ 0.195 & 11.747 $\pm$ 0.031 & 1.25 $\pm$ 0.83 & Stellar? \\
7251 & 05364890+0955423 & 84.20378 & 9.92844 & 12.817 $\pm$ 0.022 & 12.534 $\pm$ 0.023 & 12.179 $\pm$ 0.034 & 11.387 $\pm$ 0.028 & 2.61 $\pm$ 0.88 &	 \\
7615 & 05365897+0956307 & 84.24572 & 9.94187 & 14.314 $\pm$ 0.029 & 14.053 $\pm$ 0.034 & 13.940 $\pm$ 0.099 & 11.866 $\pm$ 0.038 & 1.62 $\pm$ 0.49 & Galaxy\\
\enddata
\tablecomments{}
\tablenotetext{~}{Star \#3778 (HD36861 C): is located about 30\arcsec south of the star {\LOri}. This F8 V star is member
of the central multiple system of the cluster \citep{lindroos85,bouy09}. Its strong infrared excess at 8 {\micron} 
and 24 {\micron} suggest that \#3778 is an intermediate mass star surrounded by an optically thick disk.}
\tablenotetext{~}{Star \#1959 (NLTT15297): is a double system reported as a high proper motion star. Its optical
colors reported in SIMBAD suggest that \#1959 is a foreground system. This star does not have photometry in the
optical database used in \S \ref{s:photsel}.}
\tablenotetext{~}{Star \#5968: is the only possible disk bearing star with discrepant photometric membership in 
\S \ref{s:photsel} (see text)}
\tablenotetext{~}{Star \#4244: is the faintest object with excess at 24 \micron. Since this object does not exhibit 
excess at 8\micron, \#4244 could be a transitional disk around a very low mass object in the {\LOri} cluster. 
Additional observations are required to confirm membership for this object.}
\end{deluxetable}